\documentclass[12pt,preprint]{aastex}
\slugcomment{Accepted and scheduled for publication  
in {\it The Astrophysical Journal} } 
\def\lax {\ifmmode{_<\atop^{\sim}}\else{${_<\atop^{\sim}}$}\fi}  
\def\gax {\ifmmode{_>\atop^{\sim}}\else{${_>\atop^{\sim}}$}\fi}  
\def\gtorder{\mathrel{\raise.3ex\hbox{$>$}\mkern-14mu
             \lower0.6ex\hbox{$\sim$}}}
\def\etal { et al. }

\begin{document}

\title{Determination of Black Hole Masses in Galactic Black Hole Binaries  using 
Scaling of Spectral and Variability Characteristics}

\author{Nickolai Shaposhnikov\altaffilmark{1,2} and Lev Titarchuk\altaffilmark{2,3,4} }

\altaffiltext{1}{CRESST/University of Maryland, Department of Astronomy, College Park MD, 20742, nikolai.v.shaposhnikov@nasa.gov}

\altaffiltext{2}{Goddard Space Flight Center, NASA, 
Astrophysics Science Division, code 662, Greenbelt MD 20771}

\altaffiltext{3}{George Mason University/Center for Earth
Observing and Space Research, Fairfax, VA 22030; and US Naval Research
Laboratory, Code 7655, Washington, DC 20375-5352; lev.titarchuk@nrl.navy.mil;  
Goddard Space Flight Center, NASA,  code 661, Greenbelt  
MD 20771 USA; lev@milkyway.gsfc.nasa.gov}
\altaffiltext{4}{Physics Department, University of Ferrara,
Via Saragat, 1 44100 Ferrara, Italy;  ICRANET Piazzale d. Repubblica 10-12
65122 Pescara (PE),  Italy} 

\begin{abstract}

We present a study of correlations between X-ray spectral  
and  timing properties observed  from a number of 
Galactic Black Hole (BH) binaries during   hard-soft  state spectral evolution. 
We analyze 17 transition episodes 
from 8 BH sources observed with {\it Rossi X-ray Timing Explorer}
({\it RXTE}). Our scaling technique  for BH mass determination uses  
a  correlation between spectral index and  quasi-periodic oscillation (QPO) frequency.
In addition,  we use a correlation
between index and the normalization of the disk ``seed'' component to cross-check  the BH mass determination and estimate the distance to 
 the source. While  the index-QPO correlations for two given sources  contain  information on 
the  ratio  of the BH masses in those sources,  the index-normalization correlations depend on 
the  ratio  of the BH  masses  and the  distance square ratio.
In fact, the index-normalization correlation also discloses  
the index-mass accretion rate saturation effect given that the normalization of disk ``seed'' 
photon supply is proportional to the disk mass accretion rate.
We present arguments  that this observationally established index saturation effect is 
a  signature of the bulk motion (converging) flow onto black hole which was early 
predicted by the dynamical Comptonization theory. We use GRO J1655-40
as a primary reference source for which the BH mass, distance and inclination  angle are 
evaluated  by  dynamical measurements with unprecedented precision  among other Galactic BH sources.
We apply our scaling technique to determine BH masses and distances for
Cygnus X-1, GX 339-4, 4U 1543-47, XTE J1550-564, XTE J1650-500, H 1743-322 and
XTE J1859-226. Good agreement of our results for sources with known values of BH masses 
and  distance provides an independent verification for our scaling technique. 

\end{abstract}


\section{Introduction}

Determination of masses of Galactic black holes (BHs) is 
one of the most important tasks in modern astronomy. In general, 
the knowledge of the BH mass distribution 
of the Galaxy can provide important clues
on  stellar evolution. It also can constrain the maximum mass
of a neutron star and the minimum mass of a BH. For example,  \cite{rr74} predicted that 
 the maximum mass of neutron star,  formed  as a result of a supernova explosion,
 is probably  about 3.2 solar masses.  Study of the characteristic changes in spectral 
and variability properties  of X-ray binaries proved to be a valuable source of information on the physics 
governing the accretion processes and on the fundamental parameters of BHs.

BH observational appearance is conventionally described in terms of BH state classification
\citep[see][ for different flavors of BH states definitions]{rm,bell05,kw08}.
We adopt the following general  BH state classification for 
five major BH states:{\it quiescent}, {\it low-hard} (LHS), 
{\it intermediate} (IS), {\it high-soft} (HSS) and {\it very high} 
states (VHS). When a BH transient  goes into outburst 
it leaves a quiescent state and enters LHS, a low luminosity
state with the energy spectrum dominated by a Comptonization component 
combined (convolved) with  a  weak thermal component. The photon spectrum in LHS  
is presumably a result of  Comptonization (upscattering)
of soft  photons, originated  in a relatively weak accretion disk,  off 
electrons of  the hot ambient plasma [see e.g. \cite{st80})].
Variability in LHS is high (fractional root-mean-square variability
is up to 40\%) and  presented by a flat-top broken power law (white-red noise) shape, accompanied by 
quasi-periodic oscillations (QPOs) in the range of 0.01-30 Hz, observed as narrow peaks in the power  density spectrum (PDS).
 The HSS photon spectrum is characterized by a prominent thermal 
component which is probably a signature of  a strong emission coming from a geometrically thin accretion disk.
A weak power-law component is also present at the level  
of not more than 20\% of the total source flux. In the HSS the flat-top variability 
ceases, QPOs disappear and PDS acquires a pure power-law shape.
The total  variability in HSS is usually about 5\% fractional rms.
The IS is a transitional 
stage between LHS and HSS.

Timing and spectral properties of an accreting BH are tightly
correlated (see a comprehensive set of references in Table \ref{datatab}).
Correlations between the spectral hardness (photon index) and 
the characteristic frequencies of the quasi-periodic variability
observed in the lightcurves of BH   sources has been proposed for use as a tool
to determine a BH mass \citep[][hereafter TF04]{tf04}.
\cite{ST07}, hereafter ST07, employed this idea to measure the mass of 
a prominent BH source Cygnus X-1 and obtained a BH mass of $M_{Cyg X1}=(8.7\pm 0.8)M_{\odot}$,
which has a better precision  than conventional dynamical methods.
Scaling arguments were also used to estimate masses in a number of ultra-luminous X-ray sources
 \cite{ft04,dtg06,str06}. In this Paper we further explore this technique by
analyzing more representative  data sets from Galactic BH sources.
We concentrate our efforts on the study of 
correlations between spectral index and characteristic sub-second variability frequencies  and
the total  accretion disk luminosity. These two correlation 
patterns carry the most direct information on the BH mass and   source  distance.
We also investigate the possibility that the shape of the correlation
pattern can provide a direct signature of the bulk motion (converging) flow onto  a black hole, 
or, in other words, the signature of a black hole [see \cite{tz98} and \cite{lt99} for more details on this subject].
We present  observational and theoretical arguments that
 the index-mass accretion rate saturation effect is a signature of a converging 
flow, when the mass accretion rate exceeds the Eddigton limit, which can only  exist in BH sources.

In the presented study we enhance the scaling method by considering 
joint scaling of spectral index both in 
QPO frequency and disk component normalization (disk flux) domains. In the LHS
and IS only a small part of the disk emission component is seen directly. The
energy spectrum is dominated by a Comptonized component seen as 
a power law. To calculate the total normalization
of the ``seed'' disk black body component we model the spectrum
with a Generic Comptonization model which consistently convolves 
a disk black body with the Green's  (response) function of the Compton Corona to
produce the Comptonized component (see details in \S 2). 
Calculated in this way the spectrum normalization is directly related to the BH mass, emission efficiency, 
source distance and geometry.  This allows us to apply the total disk 
normalization as an additional scaling variable and to impose an additional 
constraint on the BH mass and distance ratios. 
Thus, we use the  index-QPO and index-disk normalization 
correlations to constrain the BH masses and distances for  
Galactic BH sources H 1743-322, GX 339-4, XTE J1550-564, XTE J1650-500, 4U 1543-47, XTE J1859+226  
and Cygnus X-1 based on their expected dependences
on the mass of the central BH, the source distance and geometry.   
As our reference values for the scaling measurements, we use the 
previously measured  BH mass, system distance and inclination
for Galactic micro-quasar GRO J1655-40.
 
 The description of {\it RXTE} data for each source is given in \S \ref{data}.
We provide the details of state transitions analysis in  \S \ref{transitions} 
and we present the results of the scaling method applied to 
the data  in  \S \ref{analysis}.
We present a physical 
picture of the accretion process which is consistent with the observed phenomenology of 
state transitions  in \S \ref{theory}.   Specifically, it   provides a theoretical basis and an explanation for
the observed scaling  index-QPO and  index-normalization 
patterns.   
 in Appenix \ref{tl} we provide the mathematical details of the transition layer (TL) model  \citep[][hereafter TLM98]{tlm98}
which is the basis for the proposed scenario.
 We present  theoretical arguments 
for the inverse proportionality  of QPO frequency on BH mass in \S  \ref{inverse}.
  We discuss
the signature of bulk motion Comptonization (BMC) and its relation with
 the index evolution during 
state transition  in \S \ref{saturation}. Specifically, we show that the index saturation 
effect is a direct consequence of the existence of the innermost  bulk motion
region and, therefore, can be considered as  an observational signature 
of a BH.
In \S \ref{disc} we summarize  our results and compare 
BH masses and distances with those obtained by other 
methods. We also discuss briefly the alternative 
methods for BH mass determination from X-ray data. Conclusions
follow in \S \ref{summary}.

\section{Observations and data reduction \label{data}}

For the study presented in this Paper we have analyzed data for
{\it seventeen} spectral transitions in seven BH transient
sources observed with {\it RXTE}. We also analyzed 
a subset of Cygnus X-1 data taken during 2001-2002 years
when the source showed a transition from LHS to HSS. 
In Table \ref{datatab}
we summarize data for each transition and also provide
references to previous analyses.

We use the archival {\it RXTE} data from the HEASARC\footnote{http://heasarc.gsfc.nasa.gov/}. 
We present {\it RXTE}/PCA light\-curves  and hardness ratios  for all analyzed outbursts 
in the top panels of  Figures \ref{1655_lc}-\ref{cygx1_lc}. These diagrams are 
created using  {\it RXTE} mission long data products provided by  {\it RXTE} team.
Each data point is based on a separate pointed PCA observation. Each transition
data set is indicated by a separate color. Black color indicates observations during 
HSS or extreme LHS which are not used in our analysis.

{\it RXTE}/PCA spectra  have been extracted and analyzed in the 3.0-50 keV  energy range using the 
most recent release of PCA response calibration
(ftool {\tt pcarmf v11.1}). We  
calculated the root-mean-square normalized Fourier Power Density Spectra (PDS)
from high time-resolution PCA mode data in the frequency range 0.01-1024 Hz.
 The  relevant deadtime corrections to energy spectra 
and timing spectra have been applied following ``The RXTE Cook Book'' 
recipes and \citet{rgc00} correspondingly. The PCA energy spectra were modeled 
using XSPEC astrophysical fitting software.
Spectral analysis was done using approach similar to that adopted in ST07 for 
Cygnus X-1 and GRO J1655-40 data. To fit PCA spectra
we used the sum of the {\it bmc} component \citep[Generic Comptonization model, see][]{bmc}   and 
a Gaussian with the energy $\sim6.5$ keV, which is presumably due to 
 iron emission line. This model 
was also modified by the interstellar absorption, using the {\it wabs} model in
 XSPEC, with a hydrogen column
value fixed at the Galactic value given by nH HEASARC tool \citep{nH}
for a particular source
and a high energy cut-off ({\it highecut}). The width of the Gaussian
was constrained in the range 0.8-1.2 keV. The high energy cutoff 
component accounts for exponential overturn of the spectrum
due to electron recoil effect. Systematic error of 1.0\% have been applied 
to the analyzed spectra. 

The {\it bmc} model describes the outgoing spectrum as a convolution 
of input ``seed'' black body spectrum having
 normalization $N_{bmc}$ and color temperature $kT$
 with a Green function for Comptonization process.
Similarly to the ordinary {\it bbody} XSPEC model,
the normalization $N_{bmc}$ is a ratio of source luminosity
to square of the distance
\begin{equation}
N_{bmc}=\biggl(\frac{L}{10^{39}\mathrm{erg/s}}\biggr)\biggl(\frac{10\,\mathrm{kpc}}{d}\biggr)^2.
\end{equation}  
The resulting  model spectrum is also characterized by a parameter $\log(A)$ related
to a Comptonized fraction $f$ where $f=A/(1+A)$ and  the Green's  function spectral index
$\alpha=\Gamma-1$, where $\Gamma$ is  photon index. 

There are two reasons for using the $bmc$ model. 
 First, $bmc$ by the nature of the model is applicable to the general case when
 photons gain energy not only due to thermal Comptonization but also via
dynamic or bulk motion Comptonization \citep[see ][for details]{bmc, lt99, ST06}.
The second reason is that 
$bmc$  calculates consistently the normalization of the ``seed'' photons $N_{bmc}$,
 presumably originated in the disk.  The relation of $N_{bmc}$ and its 
 proportionality to the mass accretion rate in the disk  
is a direct consequence of the accretion disk theory \citep[see e.g. ][]{ss73}. 
The adopted spectral model decribes well the most data sets used in our analysis. 
The value of reduced $\chi^2$-statistic
$\chi^2_{red}=\chi^2/N_{dof}$, where $N_{dof}$ is the number of degrees of freedom
for a fit, is less or around 1.0 for most of the observations. For a small 
fraction (less than 3\%) of spectra with high counting statistic the value of
$\chi^2_{red}$ reaches 1.5. However, it never exceeds a rejection 
limit of 2.0.

The PDS continuum shape in LHS and IS has  usually
a band-limited noise shape, which is well presented by an 
empirical model $P_X\sim (1.0+(x/x_{*})^2)^{-in}$.
The parameter $x_*$ is related to the break frequency and
$2\times in$ is the slope of the PDS continuum after the break. 
We use Lorentzian shape to fit QPO features.
We quote the Lorentzian  centroid as a QPO frequency.

Our approach to the analysis of a BH state transition is following.
We identify a complete spectral transition as a period
when the source state  changes from LHS to HSS (or vice versa) 
showing gradual transition through IS.
In this Paper we
analyze the transitions for which the frequency of {\it RXTE}/PCA pointing observations allows to
infer the index-OPQ correlation patterns and to perform a comparative study. 
 In Table \ref{datatab} we provide the list of the transitions (outbursts)  which we 
utilized in our study.  In this Table we also introduce  an ID name 
of each transition which consists of abbreviated source name, dash, a transition direction identifier 
(``R'' for the rise and ``D'' for the decay), 
and the two last digits of the year during which the event occured. 
In Table \ref{fits_tab} we
present the details of the model  fits to the spectral and timing data
for each RXTE observation in the spectral  transition.

\section{Correlations between spectral and timing properties during
 state transitions \label{transitions}}

\subsection{Phenomenology of the correlation behavior \label{phenomen}}

In Figures \ref{1655_lc}-\ref{cygx1_lc} (top panels) we show the spectral 
evolution of a given source throughout each analyzed outburst by presenting
the PCA lightcurve and PCA hardness ratio. In the bottom panels from left to right
we plot the correlations of photon  index $\Gamma$ versus QPO frequency, {\it bmc} model
normalization $N_{bmc}$ and Comptonized fraction $f=A/(A+1)$.
Data for a particular transition are distinguished using the same color legend for 
the correlation plots and the lightcurve diagrams.

The general behavior of the correlations is following. 
The QPO frequency and normalization are positively correlated
with index. Comptonization fraction and index are usually 
anti-correlated (see right bottom panels in Figs. \ref{1655_lc}-\ref{cygx1_lc}).  
In the upper part of the correlation
patterns we observe a saturation of index which 
is more apparent in the index-normalization domain than in the index-QPO pattern.
The reason for this is that the index and normalization values,  as the best-fit parameters of the energy 
spectrum, are always 
available independently of the source state, while QPOs tend to faint as a source enters HSS.  

In ST07 we claimed the presence of index saturation in Cygnus X-1, GRO J1655-40 and GRS 1915+105 using 
correlations with QPO frequency. Here we confirm
the index saturation effect for GRO J1655-40 and Cygnus X-1
(see Fig. \ref{1655_lc} and \ref{cygx1_lc}).
Moreover, we establish  the index saturation effect   
 in GX 339-4,  H 1743-322, 4U 1543-47 and  XTE J1650-500 
(see Figs. \ref{gx339_lc}-\ref{1543_lc} and \ref{shift_figs} respectively).
It is important to mention that  the saturation of index with respect to the photon spectrum 
normalization (mass accretion rate) in XTE J1650-500
was also  recently established by \citet{mtf08} in their analysis of {\it BeppoSAX} data from
2001 outburst.

Three rise transitions, J1550-R98, H1743-R03 and J1859-R99, exhibit similar behavior. 
For all these transitions, when the  index reaches its high values ($>$2.5),  the Comptonization 
fraction shows non-monotonic behavior. Namely, for  J1550-R98 and H1743-R03,  
as the source leaves LHS $f$ 
starts to decrease as expected. However, when the source enters
 the IS  the index-$f$ track reverses and $f$ returns back to its
 maximum value of 1.0. In Figure  \ref{1743_lc}  we illustrate this evolution 
 of the Comptonization fraction $f$ as a function of photon index $\Gamma$  for the rise transition H1743-R03 
 (see red data points on the right bottom panel).  

During the decay transitions the source luminosity is lower than that for
the rise episodes and the corresponding variability is suppressed, which is probably due to 
lower  excitation power in the accretion flow \citep[see][for details]{ts08}. 
As a result  the spectral and timing data  related to
the decay transitions have lower signal-to-noise ratio
than those for the rise transitions. The index saturation levels 
for the decay stages have values close to 2.1. Most probably,
this is due to the fact that at low accretion rates  during an outburst decay
the cooling of X-ray emission  area [Comptonization cloud (CC)] by the disk photons is less 
efficient than that at high $\dot M$. As a result, the CC plasma temperature $T_e$ is higher,
leading to more efficient Comptonization and harder spectrum.
This direct connection between the temperature of the 
CC and the hardness of the emergent spectrum was shown by Monte-Carlo simulations
in \cite{lt99}.
In other words, the decay transitions appear to be different from the rise  transitions:
correlation patterns are not scalable in the normalization domain 
and they are usually not self-similar given that the index saturations take place  
at  different levels. Our experience shows 
that while decay-to-decay and rise-to-rise scaling produces 
consistent results, decay-to-rise pattern scaling is not a reliable
mass-distance indicator.

To illustrate the above points, let us consider properties of state transitions in GX 339-4.
In this source five different (rise and decay) transitions exhibited five 
different tracks (see Figure \ref{gx339_lc},  third and fourth row panels from the top). 
Two rise transitions, GX339-R02 and GX339-R07, occurred at
higher fluxes than that for two decay transitions, GX339-D03 and GX339-D07. 
These rise transitions also  showed steeper index-QPO (index-normalization) correlations with higher 
saturation levels. The difference between these 
transitions is clearly seen. On the other hand,  the GX339-R04 hard-to-soft
transition resembles a decay track with the index saturated at the value of  2.15
and a lower luminosity. But even for this rise episode the
index-normalization track is much steeper than 
those for the decay episodes. The decay transition during 2004 outburst
was quick and did not show sufficient QPO activity. We, therefore,
discarded this transition episode from our analysis.

It is also worthwhile to discuss  the behavior of XTE J1550-564 during
the 1998 rise  outburst in detail. 
The outburst started on MJD 51063 and went through the initial LHS phase,
which ended on MJD 51069. During this period spectral index increased from 1.35 to 2.2 and QPO frequency
changed from 0.3 to 3.3 Hz. 
In Fig. \ref{1550_tracks} this stage is marked by filled black circles. 
For the following 10 days the source exhibited IS with a strong flare on 
MJD 51076, when the index and QPO frequency peaked at 2.8 and 14 Hz correspondingly (IS-VHS track). 
We mark these data points by filled red circles. After the flare
the source returned to the IS with index $\sim 2.3$ and QPO at $\sim 3$ Hz.
For the next 20 days we observed smooth evolution toward the HSS
with index slightly increasing to 2.5 and QPO frequency rising to $\sim$ 5 Hz.
On about MJD 51105 the source entered the HSS (IS-HSS track). 
The  index-QPO correlation 
consists of two different tracks which are shown on Fig. \ref{1550_tracks}.
The correlation curve related to the IS-HSS stage 
is lays lower than that related to the IS-VHS transition. Clearly, the conditions in  the accretion flow
changed between these episodes. During the IS-VHS stage the source presumably went through
the strongest surge of accretion. This cold accretion flow
 provided strong cooling, which is manifested by  higher spectral index.
These changes in the physical conditions in the accretion flow led to 
a double-track pattern of the correlation. Therefore, for the scaling purposes we treat 
a combination of  the initial LHS  with IS-VHS and IS-HSS  data as two separate correlation patterns. 
We further refer to them as J1550-98Ra and J1550-98b respectively.

We revisited the data for GRO J1655-40 and Cygnus X-1, which were used for BH mass 
measurement in Cygnus X-1 (ST07). We present the data for these sources in
Figs. \ref{1655_lc} and \ref{cygx1_lc}.
In fact, Cygnus X-1 is a persistent source, with somewhat different phenomenology from
the other sources analyzed in this study. 
 The Cyg X-1 data obtained during 2001 and 2002
were  chosen for the analysis (see Fig. \ref{cygx1_lc}). 
In the 2001-2002 period the source exhibited two transition, from LHS to HSS (red data points)
 and a reverse transition marked by blue color. Black points indicate the observation
when the source was in the HSS. As seen from the lower panels, 
the rise and the decay data from Cygnus X-1 are very similar in terms of 
the index-QPO and index-normalization correlations.  In particular, the same index-normalization tracks indicate that there is no  
hysteresis effect in this source. This may be related to the persistent nature of
the accretion or the O-star wind. We, thus, use both the rise and the decay  data for Cyg X-1 data as one data set
to construct the correlation pattern. This approach is consistent with that adopted in the ST07 analysis.

\subsection{The scaling technique and parametrization of the correlation patterns \label{param}}

One of the goals of the presented study is to apply X-ray observational data 
from BH sources to infer their fundamental characteristics. To do this
we use the BH state paradigm combined with the scaling
laws expected to be observed in spectral and timing data during state
transitions. 

First, using available X-ray observations we verify a theoretical conjecture
that the  QPO low frequency is inversely proportional to a  BH mass $m=M_{bh}/M_{\odot}$ 
(see TLM98 and TF04). 
This QPO frequency dependence on   $m$ is a simple consequence  of the fact 
that all characteristic dynamical timescales in an accreting flow onto a compact object
are set by the mass of the object \citep[see e.g.][and \S \ref{inverse}]{ss73}. Thus, QPO frequencies, being
in the inverse proportion to these timescales, have to be inversely proportional 
to $m$. This intuitive law of nature now finds its observational confirmation.  
In \S \ref{inverse}  we present theoretical arguments  for  the  inverse dependence  of QPO frequencies on $m$ in terms 
of transition layer model (see details in TLM98, TF04 and Appendix A).

Thus, we write the first scaling law (that we implement in our scaling technique for 
the BH mass determination)  in terms of a scaling factor 
\begin{equation}
s_\nu=\frac{\nu_r}{\nu_t}=\frac{M_t}{M_r},
\end{equation}
where subscripts $r$ and $t$  of frequency $\nu$ denote reference and target sources respectively.

The second scaling law, which we use as a basis for our mass and distance determination analysis,
relates a source flux $F$ detected by an observer on the Earth, a disk flux  $L$ and distance $d$, namely:
\begin{equation}
\frac{F_r}{F_t}=\frac{L_r}{L_t}\, \frac{d^2_t}{d^2_r} .
\end{equation}
The disk flux $L$ can be represented as 
\begin{equation}
L=\frac{GM_{bh} \dot M}{R_*}=\eta(r_*) \dot m  L_{\rm Ed}.  
\label{lumin}
\end{equation}
Here $R_{*}=r_{*}R_{\rm S}$ is an effective radius where the main energy release takes place in the disk,  $R_{\rm S}=2GM/c^2$ is the Schwarzschild radius, $\eta=1/(2r_*)$, $\dot m=\dot M/M_{crit}$ is dimensionless mass accretion rate in units of the critical mass accretion rate
$M_{crit}=L_{\rm Ed}/c^2$ and $L_{\rm Ed}$ is the Eddington luminosity.

On the other hand 
\begin{equation}
L_{\rm Ed}=\frac{4\pi GMm_pc}{\sigma_{\rm T}}
\label{ed_lumin}
\end{equation}
i.e. $L_{Ed}\propto M$ and thus using Eqs. \ref{lumin}-\ref{ed_lumin}
we have that 
\begin{equation}
L\propto\eta(r_*) \dot m m.
\label{lumin_m}
\end{equation}

Note that both $\dot m$ and $\eta$ are assumed to be the same for two different 
sources in the same spectral state, which leads to $L_r/L_t=M_r/M_t=1/s_\nu$. 
In our spectral analysis we determine the normalization of ``seed'' 
radiation, which is supplied by an accretion flow (disk) prior to 
Comptonization. The ratio of these normalizations for
target and reference sources in the same spectral state can be written as
\begin{equation}
s_N=\frac{N_r}{N_t}=\frac{L_r}{L_t}\frac{d^2_t}{d^2_r}f_G.
\end{equation}
Here $f_G$ is a  geometrical factor which  takes into account the difference in viewing angles
of the seed photon emission area (the inner disk) by the Earth observer for two given analyzed sources.
Therefore, in the case of radiation coming directly from the disk it would have
the value $f_G=(\cos \theta)_r/(\cos \theta)_t$, where $ \theta$ is the inclination
angle of the inner disk. Note that this form of $f_G$, however, may not be very accurate
 because the geometry  of both the inner disk and the corona
 can be different from the plane (disk) geometry. In fact, in BH states of interest (LHS ans IS)  
  they can be described, for instance, 
 by quasi-spherical configuration. Despite of this uncertainty in the determination of  $f_G$, we adopt the above form $f_G$ in which  $\theta\sim i$ if the information on the system inclination 
angle $i$ is available.  
In doing this we introduce an additional assumption
that the inner accretion disk and the system inclination angles are the same. 
This assumption, however, proved to be in good agreement with our analysis.

Now we  
write  the final equations of our scaling analysis. Namely, when $s_\nu$ and $s_N$ are measured,
 the mass and the distance of the target source can be 
calculated as
\begin{equation}
M_t=s_\nu M_r
\label{mass}
\end{equation}
and
\begin{equation}
d_t=d_r\biggl(\frac{s_\nu s_N}{f_G}\biggr)^{1/2}.
\label{distance}
\end{equation}
The part of the scaling method presented by equation (\ref{distance}) is
new. In the  previous analysis (see ST07) we only use the scaling factor of the index-QPO correlation   
 $s_\nu$ and  apply it  for  determination of  the BH mass distribution.
 The new modified technique uses timing information and the normalization of photon
spectrum to constrain both BH mass and the distance for a target source,
provided that these quantities are known for a reference source.

With equations (\ref{mass}) and (\ref{distance}) in hand, the task of
BH mass and distance measurements for a target source is reduced to 
the determination of scaling coefficients $s_\nu$ and $s_N$ with
respect to the data for a reference source. The appropriate technique 
has already been implementaed in ST07. Specifically, after scalable state transition
episodes are identified for two sources, the correlation pattern for
a reference transition is parameterized in terms of the analytical function
\begin{equation}
f(x)=A-D\,B\ln\left[ \exp \left(\frac{1.0-(x/x_{tr})^\beta}{D}\right)+1\right],
\label{anat_corr}
\end{equation}
where argument  $x$ is either QPO frequency $\nu$ or normalization $N_{bmc}$.
By fitting this functional form to the correlation pattern, we find a 
set of parameters $A$, $B$, $D$, $x_{tr}$ and $\beta$ which represent
a best fit form of the function $f$ for a particular correlation curve. 
 
For $x\gg x_{tr}$ the correlation function $f(x)$ converges to constant value $A$.
Thus,  $A$ is a value of the index saturation level, $\beta$ is the power-law 
index of the  part of the curve for lower argument values
and $x_{tr}$ is the  frequency/normalization at which  index  levels off. 
Parameter $D$  controls how fast the transition occurs. 
We scale it  to a target pattern by a transform $x\rightarrow s x$ when the best fit is found for a reference pattern.

The use of  analytical function to parametrize the correlations allows  us to 
avoid theoretical modeling of the index-QPO relation based on any particular
physical model, which would require some internal parametrization in any case
(see discussion in ST07).  It  makes the scaling procedure model independent, and it is 
justified because  it satisfactory  represents both the index-QPO and the 
index-normalization patterns.

\section{BH mass and distance measurement in XRBs \label{analysis}}

The Galactic microquasar GRO J1655-40 is well studied BH X-ray binary
with the best measured mass and distance among known stellar mass BH sources.
We therefore use this system as a primary reference source for our scaling.
In addition, the ancillary information on the system inclination
is available which makes distance determination more reliable. 
However, not all sources we analyse here have transitions directly scalable to GRO J1655-40 data.
Therefore, we have to utilize intermediate reference source to provide 
measurements for all sources. Specifically, only GX 339-4 and 4U 1543-47
have transitions which are scalable to the our  primary  reference source. After
measuring BH mass and distance for GX 339-4 we use this source as a reference
for XTE J1550-564 and XTE J1650-500. Note, this  scaling between GX 339-4 
and XTE J1650-500 was allowed both with rise and decay data.
We then use  XTE J1550-564 as a reference
source for scaling to H 1743-322 and XTE J1859-226 data. Finally, we measure BH mass and distance
for Cygnus X-1 by scaling to GRO J1655-40 and 4U 1543-47. All measurements are summarized in
Table \ref{scaling_tab} and illustrated in Figures \ref{shift_figs} and \ref{shift_figs_2}.
Below we provide some details on how we combine our scalings in chains to provide
measurements for all sources of interest.

We first determine parameters for sources which have correlation patterns self-similar to
GRO J1655 data. Namely, data sets  GX399-D07 and U1543-D02 are scalable to  that of J1655-D05.
These scalings  provide BH mass measurements for GX 339-4 and 4U 1543-47 (Fig. \ref{shift_figs} panel A and B).
Other decay transitions (i.e. GX339-D03, J1650-D01 and J1743-D03)
have slightly higher saturation levels of $\sim$ 2.1 and present a different cluster of
scalable patterns (datasets). Thus we use GX339-4 as a reference source to provide 
the mass-distance determinations for XTE J1650-500 (Fig. \ref{shift_figs} panel D) 
and H 1743-322 (Fig. \ref{shift_figs} panel F).
 We  use GX339-R04 data to scale to XTE J1550-564 (J1550-R00)
data and to measure BH mass and distance in XTE J1550-564 (Fig. \ref{shift_figs} panel C).
 The resulting value of BH mass is in an
excellent agreement with the  dynamical determination of BH mass in XTE J1550-564. We also scale 
the rise transition datasets GX339-R04 and J1650-R01
(Fig. \ref{shift_figs} panel E)
to provide a double check with the decay data scaling (see Fig. \ref{shift_figs} panel D) 
for the same sources. The results obtained with these parallel scalings
 are in  a striking agreement.

Finally, we use XTE J1550-564 (J1550-R98) data 
for scaling with H 1743-322 (1743-R03) and XTE J1859-226 (J1859-R99) (Fig. \ref{shift_figs_2} 
panels G and H correspondingly).
 As we mentioned in the previous section, during the 1998 outburst rise from XTE J1550-564
the index exhibited two correlation tracks with different  saturation
levels. The initial LHS combined with subsequent IS-VHS phase
is similar to H 1743-322 transition observed during the 2003 rise episode (1743-R03).
We therefore use this subset of XTE J1550 (J1550-R98a) to scale 
to  the 1743-R03 data. The IS-HSS phase of J1550-R98 data is scalable with 
J1859-R99 transition (Fig. \ref{shift_figs_2} panel H). 
We also note that these outbursts (J1550-R98, J1859-R99 and H1743-R03) 
are similar not only in terms of the index-QPO and index-normalization behavior but also in terms of 
the index-$f$ patterns and the overall appearance of outburst lightcurves 
(we illustrate  these observational appearances in  Fig. \ref{1743_lc} which presents data for H 1743-322).

Finally, in Fig. \ref{shift_figs_2} panel K, we show the scaling between J1655-D05 and CYGX1 
datasets (see Table \ref{datatab} for these dataset descriptions). In Table \ref{scaling_tab} we present the scaling 
coefficients  $S_{\nu}$ and  $S_{N}$ obtained using our scalings procedure.
We also provide BH masses and distances obtained using the calculated  
scaling coefficients and Eqs. (\ref{mass}-\ref{distance}). 
The parameterizations of reference patterns in terms of the empirical  
functional form $f(x)$ (Eq. \ref{anat_corr}) are presented in Table \ref{partab}.

Thus, we construct 
the sequence of scalings which allows us to obtain  the BH mass and the source distance for a number 
of BH XRBs (and in some cases to provide a double check for these measurements).
Table \ref{fintab} presents the final results of
BH mass and distance determinations using our scaling method along
with independently reported masses, distances and inclination angle
values found in literature. For the sources for which we were able to make
parallel scaling measurements with two different dataset pairs, 
the final best-fit values are the averages between individual
measurements and the error ranges are calculated using the 
sum of error intervals given by individual scalings.

\section{Physical scenario for the evolution of spectral and timing 
properties during BH state transitions \label{theory}}

The accretion process onto a BH is far from
being completely understood. Theoretical models which are available for
astronomers to explain observational phenomena usually deal
with only narrow aspects of a broad observational picture.
For example, while there are several proposed QPO models in the literature 
\citep[see][for references]{rm}, most of 
them concern only QPO phenomenology and lack the connection with spectral 
properties and state transition aspects. However, in the recent years
a concept of ``transitional layer'' (TL) in the accretion flow has 
emerged. In the framework of this model several major observational
aspects of accreting BHs find natural explanation.
We argue that this concept is a strong candidate to provide a basis
for a theory of accretion process onto compact objects
which would accont for the current observational picture of BH sources.
Moreover, this paradigm may provide us with a first direct BH signature 
observationally confirmed from multiple sources.

The starting point in the development of the TL
paradigm is the notion of the necessary deviation
of the rotational profile in the innermost part of the disk 
set by the conditions at (or near) a compact object for which rotation is presumed to be much slower than 
the Keplerian frequency near the object. This problem was, the first time, considered   by TLM98 where the authors  showed  that 
the adjustment of the Keplerian rotation in  the accretion disk to the sub-Keplerian rotation of the central 
object  leads to the formation 
of the inner hot Compton corona (CC).
 
\subsection{The inverse dependence of the QPO frequency $\nu_L$ on the BH mass \label{inverse} }

TLM98 define the transition layer as a region confined between the the inner
sub-Keplerian disk boundary (where the inner TL radius $R_{in}=bR_{\rm S}$,
$ R_{\rm S}$ is the Schwarzchild radius and $b$ is a numerical factor $\lax 3$) and the first Keplerian orbit (see Figs. 1  in  TLM98 and TF04).
Numerical calculations (see also  Eq. \ref{eq:rout} and  Fig. 2 in TLM98)  showed that the 
value of dimensionless outer radius $r_{out}=R_{out}/R_{in}$ strongly depends on $\gamma-$parameter (Reynolds-number)
when the rotational velocity of central of object $\omega_0$ is 
less then the Keplerian frequency $\omega_{\mathrm{K}}$ at the TL outer boundary. 

The fact that $r_{out}$ is a function of $\gamma-$parameter  implies that the CC  
dimensionless size  $l_{cc}=  (R_{out}-R_{in})/ R_{in}= r_{out}-1$ is  a function of $\gamma$ too.
Specifically, given that $\gamma-$parameter  defines the spectral state (see TLM98 and TL04)  
{\it the CC  dimensionless size $l_{cc}$ is the same for any given spectral state of BH 
even if BH masses differ by orders of magnitude}.  Thus the CC dimensional size 
$L_{cc}=bR_{\rm S}l_{cc}(\gamma)\approx 9l_{cc}(\gamma) m$ km is proportional to
 BH mass $m=M/M_{\odot}$ for a given spectral state (or $\gamma$). 
Moreover TLM98 demonstrated that the dimensionless CC size $l_{cc}=L_{cc}/(bR_{\rm S})$ 
anti-correlates with the X-ray spectral hardness (spectral index). 
TF04 identified the QPO frequency $\nu_L$
as a normal mode oscillation frequency of the CC  near the BH.
Thus one should expect that   the QPO  frequency is inversely proportional to the BH mass
given that  the QPO frequency is proportional to  a ratio of the characteristic velocity with which the 
perturbation propagates in the CC, note it  is a function of spectral index only,  and 
the CC size $L_{cc}\propto  l_{cc}R_{\rm S}$. 
The CC plasma (sonic or magneto-acoustic) velocity  $V_{MA}$  
decreases with $\dot M$,  related to an increase of $\gamma$,   because 
in the case of the high mass accretion rate the disk provides strong soft photon 
illumination of the CC area. In other words, $V_{MA}$ is a strong function of the spectral state  (or $\gamma$).
Given that  the QPO frequency $\nu_L$ should be expressed as a ratio of $V_{MA}$ and the CC size $L_{cc}$, namely 
\begin{equation}
\nu_{L}\propto \frac{V_{MA}(\gamma)}{L_{cc}(\gamma, m)}
\approx\frac{V_{MA}(\gamma)}{9ml_{cc}(\gamma)~{\rm km}}
\label{eq:qpo_freq}
\end{equation}
one can  conclude that (i) there should be a correlation of  $\nu_L$ with  spectral state  (or $\gamma$)
and   (ii)  for a given spectral state  $\nu_L$ should be inversely proportional to BH mass $m$.
Thus, if this 
index-QPO correlation occurring  during  a  spectral transition  has a  form which is  similar
 to  that for  another BH source  then there is a  possibility to determine a ratio of BH masses for these 
 two  BH sources by sliding their correlations along $\log \nu_L$ axis  with respect to each other
(see  \S \ref{analysis} for  details of an implementation  of this scaling technique). 

Recent detailed analysis of the spectral and timing properties of X-ray emission in Cyg X-1 by \cite{ts08}
 confirms that QPO frequency $\nu_L$ indeed  correlates with the Reynolds number $\gamma$ and photon
 index $\Gamma$. 
The Comptonization $Y$ parameter (or the photon index $\Gamma$) which characterizes the efficiency of the Comptonization (see more details in Appendix B) is determined by the product of the average fractional energy change per scattering $\eta$ and mean number of scattering $N_{sc}$ in the CC.  TF04 used the results of the 
 Monte Carlo simulations of Comptonization processes in the CC and the bulk motion flow to 
demonstrate that  $\Gamma$ increases and then saturates when $\dot M$ increases.  
Thus {\it given that    $\nu_L$  and $\Gamma$ correlate  with $\dot M$  
 the QPO frequency $\nu_L$  should correlate with the photon index  $\Gamma$ as well}.
 
 \subsection{ The index saturation vs disk flux as an observational evidence of the converging flow 
 (bulk motion onto BH) \label{saturation}}
 
 As we have already pointed out, positive correlations  between photon index $\Gamma$ and QPO frequency
should be observed during the state transition, when  the corona is  cooled  down by the disk 
photons (TLM98, TF04). As the temperature of the CC drops during the transition
towards softer states, the bulk motion  Comptonization 
(BMC) effect becomes dominant in  the formation of the hard tail of X-ray spectra in BHs 
in the final spectral stage, i.e in HSS \citep[see][hereafter TZ98, and LT99]
{tz98}.  

As seen from Fig.  \ref{1655_lc}$-$\ref{cygx1_lc} the index saturation  is detected   
in index-$bmc$ normalization correlations 
for  most analyzed sources (sometimes the index  saturation is seen in  the index-QPO correlations too). 
Namely, the index saturation effect is found  in  GRO J1655, GX 339,  XTE J1550, XTE J1650,
 H 1743, 4U 1543,  Cyg X-1 (see Figs \ref{1655_lc}-\ref{1550_tracks} and  \ref{shift_figs} 
respectively).  The saturation level of index can vary from source to source.  
 Even for the same source the index saturation  can be variable for different transition episodes. 
  For example 
  GRO J1655-40  exhibits  the different index saturation levels 
 $\Gamma_{sat}\sim 2.3$, and $2.05$  during 2005 outburst rise and decay stages respectively (ST07).

This effect of index saturation vs optical depth of the bulk flow (BM) $\tau$  was first predicted
 by TZ98  and  subsequently reproduced in Monte-Carlo simulations by LT99.  
It is worth noting that the index saturation effect is an intrinsic property of the bulk motion  
onto BH  given that spectral index $\alpha=\Gamma-1$ is a reciprocal of the Comptonization parameter 
$Y$ (see  Eq. \ref{alpha_plm}) which  saturates when the BM optical depth $\tau$, or dimensionless mass accretion rate $\dot m$, increases.  
In fact, Y-parameter is a  product of  the average fractional photon energy change per scattering $\eta$ 
and the mean number of photon scatterings $N_{sc}$, i.e. $Y=\eta  N_{sc}$.

The thermal Comptonization parameter $Y\sim (4kT/m_ec^2)\tau^2$  given that in this case  
$\eta=4kT/m_ec^2$ and $N_{sc}\sim \tau^2$ for $\tau\gg 1$   \citep[see e.g.][]{rl79} 
   and, thus, the thermal Comptonization spectral index 
\begin{equation}
\alpha\sim [(4kT/m_ec^2)\tau^2]^{-1}.
\label{alpha_plmm}
\end{equation}
In the case of converging flow  
the preferable direction for upscatterred photons is the direction of bulk motion onto the BH,  i.e along its radius.
Note the fractional photon energy change is 
$$
\Delta E/E=(1-\mu_1 V_{R}/c)/(1-\mu_2 V_{R}/c).
$$ 
where $\mu_1$ and $\mu_2$  are cosines  of angles between the direction of electron velocity ${\bf n}={\bf V}_R/V_R$  and direction of incoming and outcoming (scattered) photons respectively. 

  The number of scattering of up-Comptonized photons $N_{sc}$  can be estimated as a ratio of the radial characteristic size of the 
converging flow  $L$ and the free path $l$ in the direction of motion, namely $N_{sc}\propto L/l=\tau$
given that  $\Delta E/E$ has a maximum at $\mu_2=1$ for given $\mu_1$ and $V_R$.
On the other hand  the efficiency per scattering  for bulk motion flow $\eta\propto 1/\tau$  when 
$\tau\gg 1$ \citep{lt07}.

Thus one can conclude that  {\it the Comptonization parameter $Y=\eta N_{sc}$  and  
 energy index $\alpha=Y^{-1}$} (see Eq. \ref{alpha_plm})  {\it saturate to a constant values when 
 optical depth (or mass accretion rate) of the BM flow increases}.

The index saturation value is determined 
 by the plasma temperature  during a  transition [see LT99].  On the other hand,
 (see e.g. TLM98) the plasma temperature strongly depends on the  mass accretion rate in the 
bulk motion region $\dot M_{bm}$ 
and  its illumination by the disk photons $F_{disk}$. For higher $\dot M_{bm}$ and  $F_{disk}$  the 
plasma  temperature is lower.  The level of the index  saturation decreases when  the plasma 
temperature in the bulk motion increases (TF04). 
Thus the index  saturation levels can be  different from source
 to source depending on the strength of the disk in a particular case. 
 
\section{Discussion \label{disc}}

Currently a BH identification in X-ray observational astronomy is based solely
on a mass of compact object.
Namely, a compact X-ray source is classified as a BH if
it is well established that its mass exceeds the upper mass limit of a
stable rotating neutron star, namely 3.2 M$_{\odot}$ 
\citep[see e.g.][] {rr74}.
To the date  there is only one  widely accepted  method for the mass determination related to the 
measurement of the mass function $f(M)$ based on optical spectroscopy. 
While the theoretical mass function is a function of the two masses and
the inclination angle, the value of the observationally inferred mass function
(which is the minimum possible mass of the compact object) 
depends only on the velocity half-amplitude K and the
orbital period $P$ and independent of the
inclination angle $i$ : $f(M) \propto PK^3$. 
 
\cite{o02},  has summarized the measurements of the
rotational velocities and inclinations for  17 black hole binaries.
He comments that  the difficult part  of this procedure has been related 
to the inverse problem of parameter optimization since the parameter space to search is usually 
vast and often parameters can be tightly correlated. In many cases, the observed light curve 
is not entirely due to the companion star owing to the presence of extra sources of light 
(usually from an accretion disk).
Another difficult aspect is the uncertainty in radial velocity amplitude, which
appears as third power in solution.
Therefore, while the dynamical mass measurements in a number
of BH sources were successful,  one  should always be cautious about 
potential systematic errors.

The dynamical measurement of the mass of a compact object of more
than the theoretical limit for a stable NS configuration is currently
considered  a sufficient evidence that the object is a BH. 
We note, however, that  by itself such a mass measurement is not a direct proof of the nature
of an object as a BH. Such a proof, for example, would come from the confirmation 
of the presence of a BH horizon. Direct observations of a BH horizon, 
by theory are not possible. Therefore, one
has to look for signatures which would require the BH horizon
and therefore would serve as a proof of its presence. In this work
we present theoretical arguments and observational evidence 
in support that the index saturation effect observed during BH spectral transitions 
is due to the dynamical Comptonization in the
converging bulk motion  in the innermost part of a compact
object. Formation of this bulk motion region is not possible in the presence of 
any solid surface, which would otherwise be manifested by  observations of a ``feed-back''
in the form of a strong spectral component due to the energy release on the 
surface and/or coherent pulsations as in case of neutron stars.
In fact, none of this ``feed-back'' features were ever observed from these
sources. Therefore, we argue, that the photon index saturation with the mass 
accretion rate is the signature of the bulk motion (converging) flow onto BH.
In this Paper we present the observational evidence of the index saturation  detected in several
X-ray outbursts from BH sources.

We measure the BH mass and distance 
using scaling relationships in the correlations between  spectral and
timing properties observed during the state transitions in Galactic
accreting BHs. We complement previously proposed technique for 
BH mass determination using index-QPO shift (ST07)  with scaling the index$-$X-ray spectrum normalization  which contains information 
on a source distance. Using our modified scaling approach we
calculated BH mass and distances for seven sources. Results of these measurements
are given in Table \ref{fintab}. We also present constraints on BH masses and 
distances which are available using other methods. 

First, we compare the BH masses found using dynamical
methods with masses derived by the scaling technique. BH masses provided by
scaling are in  good agreement with those given  by the dynamical method
for  4U 1543-47 and XTE J1550-564.  For 
XTE J1859-226 and Cygnus X-1,   BH mass constraints  based on optical   
 measurements   are consistent with BH mass values obtained using the scaling technique.
The BH mass of $12.3\pm1.4 M_\odot$ obtained for GX 339-agreement 
with the lower limit of $6 M_\odot$ provided by \citet{munos08}. 
For H 1743-322 mass measurement by the dynamical method is not available.
Based on the values of high frequency QPOs \citet{mcc07} concluded that
the H 1743-322 has a high inclination ($i>70^\circ$) and the BH mass about 
11 solar masses. Our  BH mass  values of $m=13.3\pm3.2$ 
 agrees with this estimate.
 
For XTE J1650-500 we obtain a new value for BH mass of $m=9.7\pm1.6$,
which is considerably higher than the  the upper limit of 7.3 $M_\odot$ inferred  by \citet{oro04}.
In this paper we also retract our previous claim of the measurement of
BH mass of $3.8\pm0.5$ in XTE J1650-500 \citep{st08}. 
This preliminary claim for XTE J1650-500 was
based on self-similarity of XTE J1650-500 and  XTE J1550-564 index-QPO data only.
In our study of index correlations
with spectrum normalization and other parameters it has become clear 
that the data sets used to obtain low mass value for XTE J1650-500 (J1650-R01 and J1550-98a) 
are not scalable. The new value of BH mass for XTE J1650-500 is 
obtained by scaling with GX339-4 using more careful consideration of  scalability
criteria.  It is also worth noting that  \citet{oro04} 
remarked that their analysis were affected by poor statistic 
and limited number of templates. Moreover, their conclusion on
the BH mass upper limit was based on the number of assumptions about the spectral class of the
optical companion, accretion disk contribution and inclination angle. 

We hope that future observations will clarify the discrepancy in XTE J1650-500 mass
determinations by different methods. We, however, note an interesting fact
that most our inferred BH masses, as well as majority of BH masses measured
by dynamical methods fall in a narrow range around 10 $M_{\odot}$. This may be due 
to some observational selection effect or some currently 
unknown factors in stellar evolution. In our opinion,  this matter deserves a 
 separate investigation.


In most cases our distance estimates are also  in agreement with 
distances given  by independent methods. Especially encouraging is a
 good agreement for 4U 1543-47 and Cygnus X-1. For this sources 
the inclinations are much smaller than that for the reference source 
GRO J1655-40. This means that an approximation of geometrical factor 
$f_G$ as a ratio  of cosines of inclination angles works well.  
For H 1743-47 \citet{mcc07} inferred the inclination angle of $i\sim 70^\circ$ and the distance of 10 kpc.
Using this  inferred value of $i$  we obtained  that $f_G\sim 1$, 
and the distance of $d=9.1\pm1.5$ kpc  for  H 1743-322.


An X-ray spectroscopic  method  of the BH mass determination  was implemented
by \cite{b99} and further developed in \cite{st99}.
The method is applicable to the HSS data when the thermal blackbody-like (BB) component, which is presumably related to the accretion disk
emission, dominates
in the emergent  X-ray spectrum.
The main idea of this X-ray spectroscopic method  is 
to use the color temperature and  normalization of the BB component  to infer the BH mass. 
One needs to employ a correct color factor $T_h$ which is a 
ratio of the color and effective  BB temperatures of the disk. 
For the color factor calibration  \cite{b99}  chose  GRO J1655-40 for which   $d$,  $m$ and $i$  
are well determined. Analyzing the {\it RXTE} spectra of  this source and using 
the Shakura-Sunyaev (SS) disk  model \citep{ss73} they obtained 
$T_h\sim 2.6$. This is drastically  different from the commonly used value of 1.7, obtained by 
 \cite{st95} also in the framework SS model but with certain assumptions regarding the disk viscosity 
parameter. 
Thus \cite{b99} and later \cite{st99} used GRO J1655-40 calibrated  value of the color ratio of $2.6$ for 
the BH mass determination.
The latest results of the BH mass determination for a number of the BHC sources  using 
the X-ray spectroscopy method  are summarized in \cite{st03}, hereafter ST03. We also present these values
in Table \ref{fintab} for comparison. 
From this comparison
one may see that the main uncertainties of the ST03 BH mass estimates, which relied on
the spectroscopic method, 
are mostly  driven by uncertainties of the inclination $i$, the distance $d$ and 
by the value of the color 
factor $T_h$. 

\section{Conclusions \label{summary}} 

We perform a comprehensive study of a highly representative 
set of well observed spectral transitions in Galactic BH sources.
The goal of the study is to further explore the possibility 
of measuring the BH fundamental properties from X-ray 
data and search for BH signatures. We use correlations between spectral and timing 
properties during the state transitions as a main tool to investigate
the BH spectral transition phenomenology. 

We compiled the state transition data from eight galactic BH sources 
collected with the {\it RXTE} mission. We examine the correlation between photon 
index of the Comptonized spectral component, its normalization and the
QPO frequency. Analyzing  the behavior of the correlation patterns
we utilize four basic scaling laws: the inverse proportionality of 
 frequencies of oscillatory processes to the mass of central BH
and the disk flux  proportionality  to the BH mass,  mass accretion rate and its inverse proportionality  to  a square  of distance. We establish
that scalable correlation patterns indeed contain information on 
BH mass and system distance.

We then combine the  scaling of the correlation patterns 
in the frequency and normalization domains  for a set of Galactic X-ray BH binaries 
from which we determine BH masses and distances.
Our  results confirm that the correlation 
scaling method 
is a powerful technique for a BH mass determination.
The application of the scaling technique 
for the high precision measurements of BH masses requires {\it very well sampled observations of
the source evolution through the outbursts and careful consideration of scalability 
of correlation patterns}.



We have tested the scaling method using the known (from optical, IR observations and 
X-ray spectroscopic measurement) BH masses in  4U 1543-47, XTE J1859+226,
XTE J1550-564 and Cygnus X-1. 
Using the inverse proportionality of QPO frequency  to a BH mass for a given spectral state 
we arrive to a set of BH mass determinations, which are in  good agreement
with available dynamical data for  several systems. 

We also show  observationally  that the transition layer theory (TLM98, TF04)
correctly predicts 
the  QPO frequency dependence on the BH mass
and overall shape of the correlations between spectral and timing
properties observed during spectral transitions.
{\it The success of the scaling method  for the BH  mass determination 
strongly supports the  Compton Cloud (CC) origin of the observable index-QPO correlation.}

We present the observational evidence supporting
the theory of the bulk motion (converging) flow,  i.e.  the index saturation with the mass accretion rate. 
Specifically, analyzing {\it RXTE} observations for a set of BH outbursts we find
 that the index saturation is seen   in the index-QPO and index-normalization correlations.
We  argue that the index saturation with mass accretion rate as a signature of the   bulk  (converging) 
flow should  only exist in the BH sources. 
Only in these sources 
there is no radiation pressure feedback effect at high mass accretion rate, as that takes place in  NS binaries.
 In other words, {\it this index saturation effect provides  a robust observational 
evidence for the presence of black holes in these BHCs.}

The authors acknowledge the referee for many constructive suggestions to improve the paper quality and its presentation. 

\appendix

\section{The transition layer paradigm \label{tl}}
The radial motion in the disk is controlled by friction and the angular
momentum exchange between adjacent layers, resulting in the loss of the
initial angular momentum by accreting matter. The corresponding radial
transport of the angular momentum in a disk is described by the following
equation [see e.g. \cite{ss73}]: 
\begin{equation}
\dot{M}\frac{d}{dR}(\omega R^{2})=2\pi \frac{d}{dR}(W_{r\varphi }R^{2}),
\label{eq:mot}
\end{equation}%
where $\dot{M}$ is the accretion rate in the disk and $W_{r\varphi }$ is the
component of a viscous stress tensor that is related to the gradient of the
rotational frequency $\omega =2\pi \nu $, namely, 
\begin{equation}
W_{r\varphi }=-2\eta HR\frac{d\omega }{dR},  \label{eq:tens}
\end{equation}%
where $H$ is a half-thickness of a disk and $\eta $ is turbulent viscosity.
The non-dimensional parameter that is essential for equation (\ref{eq:mot})
is the Reynolds number for the accretion flow, 
\begin{equation}
\gamma =\dot{M}/4\pi \eta H=RV_{r}/D,  \label{eq:rey}
\end{equation}%
where $V_{r}$ is a characteristic velocity, and $D$ is the diffusion
coefficient. $D$ can be defined as $D=V_{t}l_{t}/3$ using the turbulent
velocity and the related turbulent scale, respectively or as $%
D=D_{M}=c^{2}/\sigma $ for the magnetic case where $\sigma $ is the
conductivity [e.g. see details of the D-definition in \cite{lang98}].  

Equations $\omega =\omega _{0}$ at $R=R_{in}=bR_{\rm S}$ (at the inner TL radius), 
$\omega =\omega _{\mathrm{K}}$ at $R=R_{out}$ [the radius where the
transition layer adjusts to the Keplerian motion for which $\omega_{\mathrm{K}}= (GM/R^3)^{1/2}$], 
and $d\omega /dR=d\omega _{\mathrm{K}}/dR$ at $R=R_{out}$ were assumed by TLM98 to be the boundary 
conditions. Note that here we set the inner boundary at  $R_{in} =bR_{\rm S}$  with  $b$ of 3. This 
value of $R_{in}$ is  valid for a BH whose spin  $a$ is less than 0.8.
 
 Thus, the profile $\omega (R)$ and the outer radius of the
transition layer are uniquely determined by the boundary conditions and the
angular momentum equation (\ref{eq:mot}-\ref{eq:tens}) for a given value of
the Reynolds number $\gamma $ (see Eq. \ref{eq:rey}). 

The solution of angular momentum equation (\ref{eq:mot}-\ref{eq:tens})  satisfying 
the above boundary conditions is equation (10) in TLM98 while the following  equation  (see  also Eq. 11 in TLM98) 
\begin{equation}
3\theta _{out}/2=D_{1}\gamma r_{out}^{-\gamma }+2(1-D_{1})r_{out}^{-2}
\label{eq:rout}
\end{equation}
determines $r_{out}=R_{out}/R_{in}$ as a function of $\gamma -$parameter.
Here $\theta _{out}=\omega _{k}(r_{out})/\omega _{0}$ and $D_{1}=(\theta
_{out}-r_{out}^{-2})/(r_{out}^{-\gamma }-r_{out}^{-2})$. 

The adjustment of
the Keplerian disk to the sub-Keplerian inner boundary creates conditions
favorable for the formation of a hot plasma outflow at the outer boundary of
the transition layer (TLM98), because the Keplerian motion (if it is
followed by sub-Keplerian motion) must pass through a super-Keplerian
centrifugal barrier region. 

Note one can see that the  right hand side of Eq. (\ref{eq:rout}) is linearly  proportional to  
$\theta_{out}$ if $\theta_{out}\gg1$ because $D_1\propto\theta_{out}$ and $|D_1/r_{out}^2|\gg 1$   for $\theta_{out}\gg1$ and 
$\gamma>2$.  As a result $\theta_{out}$ can be canceled from left and right hand sides of Eq. (\ref{eq:rout})  when   $\theta_{out}\gg1$.  Namely the value of  dimensionless outer radius  $r_{out}$  strongly depends on $\gamma-$parameter ($Re-$number)   and independent of $\theta_{out}$ for $\theta_{out}>>1$. 

 It implies  that the CC  dimensionless size 
$l_{cc}=  (R_{out}-R_{in})/ R_{in}= r_{out}-1$ is a function $\gamma-$parameter ($Re-$number)  only if    $\theta_{out}\gg1$. 
It is worth noting that in the general case 
the size $l_{cc}$ is a function of $\gamma$ and  $\omega_0$ (or  BH spin $a$) too.
The direct scaling of $L_{cc}=bR_{\rm S}l_{cc}(\gamma, a)$ with BH mass  $m$ is not possible anymore.  
There is a systematic shift of the values of $l_{cc}(\gamma, a)$ for a given $\gamma$ because of BH spin $a$.

 \section{Spectral index $\alpha$ as a reciprocal of the Comptonization parameter $Y$}

The intensity of the injected soft photons  undergoing  $k$ scatterings in the Compton cloud is
\begin{equation}
I_k\propto p^k
\label{int_k}
\end{equation}
where $p$ is a mean probability of photon scattering in the CC. 
The probability of  photon scattering is directly related to the mean number of scatterings 
\begin{equation}
N_{sc}=\sum_{k=1}^{\infty}kp^kq=p/(1-p)
\label{scat_number}
\end{equation}
where $q=1-p$ is a probability of the photon escape from the CC.
Thus, using Eq. (\ref{scat_number}), we obtain 
\begin{equation}
p=1-1/(N_{sc}+1).
\label{p_vs_sc_number}
\end{equation}
Because the average photon energy change   per scattering  $<\Delta E> =\eta E$ (where $\eta>0$ for upscattering case),  the injected photon energy after $k$ scatterings  $E$ is
\begin{equation}
E=(1+\eta)^{k}E_0
\label{E_k}
\end{equation} 
The combination of Eqs. (\ref{int_k}),  (\ref{E_k})  yields that the emergent upscattering spectrum of the soft photon of energy $E_0$ in the bounded Compton cloud is a power law
\begin{equation}
I_{E}\propto\left(\frac{E}{E_0}\right)^{-\alpha}
\label{power_law}
\end{equation}
 which energy index is
\begin{equation}
\alpha=\frac{\ln(1/p)}{\ln(1+\eta)}.
\label{alpha_pl}
\end{equation}
 Using Eq. (\ref{p_vs_sc_number})
  we can reduce Eq. (\ref{alpha_pl})  to 
\begin{equation}
\alpha\approx(\eta N_{sc})^{-1}=Y^{-1}.
\label{alpha_plm}
\end{equation}
for $N_{sc}\gg1$ and $\eta\ll 1$.

\begin{deluxetable}{llllll}
\tablewidth{0pt}
\tablecaption{Spectral Transitions Data Used in Analysis}
\tablehead{\colhead{Source}&\colhead{Transition ID} & \colhead{Start}& \colhead{End}&\colhead{Type}&\colhead{Refs.}}
\startdata
GRO J1655-40 & J1655-R05 & 22/02/05 & 13/03/05 & RISE  & 11 \\
             & J1655-D05 & 11/09/05 & 27/09/05 & DECAY & 11 \\
\hline
GX 339-4     & GX339-R02 & 23/04/02 & 18/05/02 & RISE  & 4,5 \\
             & GX339-D03 & 20/02/03 & 05/06/03 & DECAY & \\
             & GX339-R04 & 10/07/04 & 20/08/04 & RISE  & \\
             & GX339-R07 & 22/01/07 & 18/02/07 & RISE  & \\
             & GX339-D07 & 07/05/07 & 01/06/07 & DECAY & \\
\hline
4U 1543-47   & U1543-D02 & 14/07/02 & 31/07/02 & DECAY & 9,10 \\
\hline
XTE J1550-564& J1550-R98 & 08/09/98 & 16/10/98 & RISE  & 1,2,3 \\
             & J1550-R00 & 10/04/00 & 05/06/00 & RISE  & 1,2,3 \\
             & J1550-D00 & 05/06/00 & 16/07/00 & DECAY & 1,2,3 \\

\hline
XTE J1650-500& J1650-R01 & 06/09/01 & 28/09/01 & RISE  & 7 \\
 	     & J1650-D01 & 19/11/01 & 25/11/01 & DECAY & 7 \\
\hline
H 1743-322   & H1743-R03 & 28/03/03 & 27/04/03 & RISE  & 3 \\
             & H1743-D03 & 15/10/03 & 03/11/03 & DECAY &  \\
\hline
XTE J1859+226& J1859-R99 & 09/10/99 & 27/10/99 & RISE  & 6 \\
\hline
Cygnus X-1   & CYGX1     & 15/06/01 & 01/12/02 & MIXED & \\
\enddata
\tablerefs{$^1$\citet{rod03},$^2$\citet{rod04},$^3$\citet{mcc07},$^4$\citet{bel06},$^5$\citet{bel05},
$^6$\citet{cas04},$^7$\citet{rossi04},$^8$\citet{tru01},$^9$\citet{kal05},$^{10}$\citet{park04},$^{11}$ST07,$^{12}$\citet{trud01},$^{13}$\citet{vig}}
\label{datatab}
\end{deluxetable}

\begin{deluxetable}{lllllllllllll}
\rotate
\tablewidth{0pt}
\tabletypesize{\scriptsize}
\tablecaption{Spectral and Timing Characteristics for {\it RXTE} data\tablenotemark{a}}
\tablehead{
\colhead{Transition ID} &
\colhead{Observation ID} &
\colhead{MJD\tablenotemark{b}} &
\colhead{BH State} &
\colhead{$\nu_{QPO}$} &
\colhead{$\alpha$} &
\colhead{$kT$} &
\colhead{$log(A)$} &
\colhead{$N_{BMC}$} &
\colhead{$E_{cut}$} &
\colhead{$E_{fold}$} &
\colhead{$F_{X}\times10^8$} &
\colhead{$\chi^2_{red}$}\\
\colhead{} &
\colhead{} &
\colhead{day} &
\colhead{} &
\colhead{Hz} &
\colhead{} &
\colhead{keV} &
\colhead{} &
\colhead{$L_{39}/d^2_{10}$} &
\colhead{keV} &
\colhead{keV} &
\colhead{erg/s/cm$^2$} &
\colhead{}
}
\startdata
GX339-R07 & 92428-01-02-00 & 54122.27 &LHS& - &    0.455(9) &     0.61(3) &     0.39(3) &   0.0566(8) &       26(3) &       76(14) &    0.872(2) &     0.82 \\
 & 92428-01-03-00 & 54127.11 &LHS& - &     0.50(2) &     0.55(5) &     0.33(6) &    0.070(2) &       28(6) &       50(29) &    1.089(9) &     0.93 \\
 & 92035-01-01-00 & 54129.47 &LHS&    0.164(6) &     0.51(1) &     0.65(3) &     0.44(2) &    0.076(1) &       21(2) &       95(9) &    1.160(2) &     0.73 \\
 & 92035-01-01-01 & 54128.94 &LHS&    0.142(3) &    0.484(8) &     0.53(2) &     0.19(3) &    0.080(1) &       26(1) &       66(6) &    1.137(3) &     0.87 \\
 & 92035-01-01-02 & 54131.10 &LHS&    0.180(4) &    0.527(9) &     0.59(3) &     0.38(3) &   0.0834(9) &       26(1) &       67(6) &    1.276(3) &     1.03 \\
 & 92035-01-01-03 & 54130.13 &LHS&    0.167(4) &     0.48(1) &     0.55(2) &     0.20(3) &    0.086(1) &       22(1) &       73(6) &    1.214(3) &     0.86 \\
 & 92035-01-01-04 & 54132.08 &LHS&    0.200(3) &     0.53(1) &     0.59(3) &     0.39(3) &    0.085(1) &       24(1) &       71(6) &    1.314(2) &     1.03 \\
 & 92035-01-02-00 & 54133.00 &LHS&    0.237(8) &     0.55(1) &     0.58(3) &     0.39(3) &    0.088(1) &       25(4) &       72(19) &    1.354(3) &     0.90 \\
 & 92035-01-02-01 & 54133.92 &LHS&    0.264(6) &    0.554(9) &     0.59(3) &     0.37(3) &   0.0899(9) &     25.2(9) &       68(6) &    1.362(3) &     0.98 \\
 & 92035-01-02-02 & 54135.03 &LHS&    0.296(4) &    0.569(9) &     0.57(3) &     0.38(3) &    0.090(1) &     24.5(8) &       60(4) &    1.380(3) &     0.83 \\
 & 92035-01-02-03 & 54136.02 &LHS&    0.361(6) &     0.56(1) &     0.60(3) &     0.37(3) &    0.094(1) &     22.7(9) &       72(5) &    1.414(3) &     1.16 \\
 & 92035-01-02-04 & 54137.00 &LHS&    0.432(5) &     0.60(1) &     0.62(3) &     0.43(2) &    0.100(1) &     22.8(8) &       70(4) &    1.510(3) &     0.76 \\
 & 92035-01-02-08 & 54137.85 &LHS&     0.53(2) &     0.63(1) &     0.54(3) &     0.33(4) &    0.104(2) &       21(1) &       78(8) &    1.522(3) &     0.98 \\
 & 92035-01-02-07 & 54138.83 &LHS&    0.900(5) &     0.64(1) &     0.52(2) &     0.16(3) &    0.117(2) &     21.6(7) &       68(3) &    1.521(3) &     0.78 \\
 & 92035-01-02-06 & 54139.94 &IS&    0.987(5) &     0.79(1) &     0.61(2) &     0.36(2) &    0.111(1) &     21.9(7) &       65(3) &    1.466(2) &     0.73 \\
 & 92035-01-03-00 & 54140.20 &IS&    1.136(2) &     0.82(1) &     0.56(2) &     0.25(3) &    0.117(2) &     21.0(8) &       75(5) &    1.438(2) &     0.72 \\
 & 92035-01-03-01 & 54141.05 &IS&    1.691(3) &     0.92(1) &     0.55(2) &     0.15(2) &    0.131(2) &     20.6(9) &       79(6) &    1.384(3) &     0.74 \\
 & 92035-01-03-02 & 54142.04 &IS&    2.434(9) &     1.03(1) &     0.51(1) &    -0.14(2) &    0.182(5) &     20.6(8) &       72(5) &    1.343(2) &     0.85 \\
 & 92035-01-03-03 & 54143.02 &IS&    3.508(9) &     1.22(1) &     0.55(1) &    -0.17(1) &    0.208(4) &       24(1) &       67(10) &    1.338(2) &     0.91 \\
 & 92428-01-04-00 & 54143.87 &IS&     4.34(1) &     1.29(1) &    0.597(9) &    -0.25(1) &    0.226(3) &       22(1) &       80(11) &    1.337(3) &     1.00 \\
 & 92428-01-04-01 & 54143.95 &IS&     4.24(1) &     1.26(1) &     0.52(1) &    -0.35(1) &    0.265(7) &       22(1) &       87(14) &    1.336(3) &     1.08 \\
 & 92428-01-04-02 & 54144.09 &IS&     4.13(2) &     1.29(1) &     0.57(1) &    -0.23(1) &    0.230(5) &       22(3) &      125(46) &    1.361(3) &     0.97 \\
 & 92428-01-04-03 & 54144.87 &IS&     5.00(2) &     1.35(1) &     0.59(1) &    -0.33(1) &    0.253(5) &       25(3) &       81(33) &    1.346(4) &     1.31 \\
 & 92428-01-04-04 & 54145.96 &IS&     5.63(2) &     1.39(1) &    0.597(9) &    -0.38(1) &    0.270(5) &       22(2) &       89(33) &    1.346(3) &     1.17 \\
 & 92035-01-03-05 & 54145.11 &IS&     5.80(2) &     1.35(1) &    0.554(8) &    -0.50(1) &    0.307(6) &       25(2) &       81(17) &    1.303(3) &     1.37 \\
 & 92035-01-03-06 & 54146.03 &IS&     5.55(2) &     1.36(1) &    0.601(8) &    -0.40(1) &    0.269(4) &       21(2) &      111(24) &    1.343(2) &     1.29 \\
 & 92035-01-04-00 & 54147.01 &IS&     6.67(2) &     1.53(1) &    0.626(7) &   -0.608(9) &    0.338(5) & - & - &    1.371(3) &     1.59 \\
 & 92035-01-04-01 & 54148.14 &IS& - &     1.26(3) &    0.796(7) &    -0.96(1) &    0.194(5) & - & - &    1.024(6) &     1.07 \\
 & 92035-01-04-02 & 54149.69 &HSS& - &     1.19(2) &    0.782(4) &   -1.104(8) &    0.193(4) & - & - &    0.963(6) &     1.27 \\
\enddata
\tablenotetext{a}{Table in the paper/PDF version includes only data for GX339-R07 transition. Entire table is available in the electronic edition of the Paper.}
\tablenotetext{b}{Date at the start of the RXTE observation}
\label{fits_tab}
\end{deluxetable}

\newpage

\begin{deluxetable}{llclllllll}
\tabletypesize{\scriptsize}
\tablewidth{0pt}
\tablecaption{Index-QPO and Index-Norm Scaling Coefficients}
\tablehead{
\colhead{Target Tr. ID}&
\colhead{Reference Tr. ID}&
\colhead{Fig. \ref{shift_figs}}&
\colhead{$S_\nu$}&
\colhead{$S_N$}&
\colhead{$m$}&
\colhead{$d$}&
\colhead{$f_G$}
\\
\colhead{}&
\colhead{}&
\colhead{Panel}&
&&
\colhead{}&
\colhead{}&
}
\startdata
GX339-D07  & J1655-D05  & A  & 1.96$\pm$0.12   & 1.65$\pm$0.06  & 
12.3$\pm$1.4   & 5.75$\pm$0.64 & 1\\
GX339-R04  & J1655-R05  &    & 1.95$\pm$0.19   & 1.9$\pm$0.1   & 
12.3$\pm$1.8   & 6.15$\pm$0.83 & 1 \\
U1543-D03  & J1655-D05  & B  & 1.49$\pm$0.15   & 2.23$\pm$0.22  & 
 9.4$\pm$1.4   & 3.5$\pm$0.5 &0.37 \\
J1550-R00  & GX339-R04  & C  & 0.87$\pm$0.02   & 0.37$\pm$0.02 & 
10.7$\pm$1.5   & 3.3$\pm$0.5  & 1\\
J1650-D01  & GX339-D03  & D  & 0.80$\pm$0.02   & 0.47$\pm$0.01  & 
9.9$\pm$1.4   & 3.5$\pm$0.5  &1\\
J1650-R01  & GX339-R04  & E  & 0.76$\pm$0.02   & 0.358$\pm$0.004 &
9.3$\pm$1.2   & 2.95$\pm$0.4 & 1\\
H1743-D03  & GX339-D03  & F  & 1.15$\pm$0.07   & 2.16$\pm$0.13  & 
 12.7$\pm$2.6   & 9.1$\pm$2.0 & 1\\
H1743-R03  & J1550-R98a  & G  & 1.19$\pm$0.08   & 6.44$\pm$0.52  & 
14.2$\pm$2.4   & 9.1$\pm$1.6 & 1\\
J1859-R99  & J1550-R98b  & H  & 0.696$\pm$0.015 & 0.429$\pm$0.005 & 
7.7$\pm$1.2   & 4.2$\pm$0.5 &1\\
CYGX1      & J1655-D05  & K  & 1.17$\pm$0.03   & 0.144$\pm$0.003 &
7.4$\pm$0.6   & 2.04$\pm$0.18 & 0.42\\
CYGX1      & U1543-D03  &    & 0.85$\pm$0.02   & 0.064$\pm$0.001 &
8.0$\pm$0.9   & 2.1$\pm$0.4 & 0.42\\
\enddata
\tablenotetext{a}{The column contains estimates for $f_G$ based on available
information on inclination angles.}
\label{scaling_tab}
\end{deluxetable}


\begin{deluxetable}{llllllllll}
\tabletypesize{\footnotesize}
\tablewidth{0pt}
\tablecaption{Parametrizations for reference patterns }
\tablehead{
\colhead{Transition}&
\multicolumn{4}{c}{Index-QPO}& 
\multicolumn{5}{c}{Index-Normalization}\\
&
\colhead{B}&
\colhead{$\nu_{tr}$}&
\colhead{D}&
\colhead{$\beta$}&
\colhead{A}&
\colhead{B}&
\colhead{$N_{tr}$}&
\colhead{D}&
\colhead{$\beta$}
}
\startdata
J1655-R05 & 0.59$\pm$0.04&3.0$\pm$0.1 & 1.0 & 1.6$\pm$ 0.2 & 2.2& 0.400$\pm$0.003  & 0.061$\pm$0.003& 1.0 & 2.3$\pm$0.6\\
J1655-D05 & 0.58$\pm$0.01 & 10.9$\pm$0.4 & 0.1 & 1.0 & 2.02$\pm$ 0.02 & 0.44$\pm$0.02 & 0.026$\pm$0.002 & 1.0 & 1.88$\pm$0.25\\
GX339-D03 & 0.66$\pm$0.02 & 5.4$\pm$0.4 & 0.1 & 1.0 & 2.08$\pm$0.01 & 0.45$\pm$0.02 & 0.010$\pm$0.001 & 1.0 & 2.0$\pm$0.3 \\
GX339-R04 & 0.58$\pm$0.02 & 1.4$\pm$0.2 & 1.0 & 1.0 & 2.14$\pm$0.01 & 0.51$\pm$0.02 & 0.039$\pm$0.002 & 1.0 & 3.5 \\
J1550-R98a & 1.27$\pm$0.02 & 1.84$\pm$0.07 & 1.0 & 0.65$\pm$0.02 & 2.94$\pm$0.08 & 1.8$\pm$0.3 & 0.162$\pm$0.05 & 1.0 & 0.6$\pm$0.1 \\
J1550-R98b & 1.32$\pm$0.03 & 1.97$\pm$0.07 & 1.0 & 0.5 & 2.55 & 1.25$\pm$0.04 & 0.164$\pm$0.004 & 1.0 & 1.0 \\
\enddata
\label{partab}
\end{deluxetable}


\begin{deluxetable}{lllllllll}
\tabletypesize{\scriptsize}
\tablewidth{0pt}
\tablecaption{BH Masses and Distances}
\tablehead{
\colhead{Source}&
\colhead{$M_{dyn}\tablenotemark{a},$}&
\colhead{$i,$\tablenotemark{a}}&
\colhead{$d$,\tablenotemark{b}}&
\colhead{$M_{ST03}\tablenotemark{c}$}&
\colhead{$M_{scal}$}&
\colhead{$d_{scal}$,}&
\colhead{}&
\colhead{Refs.}\\
&
\colhead{$M_\odot$}&
\colhead{deg}&
\colhead{kpc}&
\colhead{$M_\odot$}&
\colhead{$M_\odot$}&
\colhead{kpc}&
&
}
\startdata
GRO J1655-40\tablenotemark{d} &  6.3$\pm$0.3& 70$\pm$1 & 3.2$\pm$0.2 & & & &  & 1,2 \\
GX 339-4  & $>6$ & - & 7.5$\pm$1.6 & 9.0$\pm$3.8(4) & 12.3$\pm$1.4 &   5.75$\pm$0.8 & &5,6\\
4U 1543-47  & 9.4$\pm$1.0 &20.7$\pm$ 1.5& 7.5$\pm$1.0 & 14.8$\pm$1.6 (6.4) & 9.4$\pm$1.4 & 9.4$\pm$1.8 &&7,8\\
XTE J1550-564  & 9.5$\pm$1.1 &72$\pm$5& $\sim$2.5,$\sim$6 &  9.4$\pm$2.1 (5) & 10.7$\pm$1.5 & 3.3$\pm$0.5 &  &9,10,11\\
XTE J1650-500  &$2.7-7.3$ & $>50$ & 2.6$\pm$0.7 &  10.6$\pm$4.0 (5) &  9.7$\pm$1.6 & 3.3$\pm$0.7 & &12,13\\
H 1743-322 & $\sim$11 & $\sim 70$ & $\sim$10 & - & 13.3$\pm$3.2 & 9.1$\pm$1.5 & &14\\
XTE J1859-226  &7.6 - 12.0& - & 11 &  12.3$\pm$1.7 (11) & 7.7$\pm$1.3 & 4.2$\pm$0.5 & &15,16\\
Cygnus X-1  & 6.8 - 13.3&35$\pm$ 5&2.5$\pm$0.3  & - & 7.9$\pm$1.0 & 2.2$\pm$0.3 & &3,4\\
\enddata
\tablenotetext{a}{Dynamically determined BH mass and system inclination}
\tablenotetext{b}{Source distances found in literature}
\tablenotetext{c}{$M_i=M \cos_i$. Source distances used by ST03 are given in paranthesis.}
\tablenotetext{d}{GRO J1655-40 is a primary reference source. All masses and distances given in columns 9 and 10 are determined with respect to the best measured parameters for this source.}
\tablerefs{$^1$\citet{greene01},$^2$\citet{hr95},$^3$\citet{her95},$^4$\citet{ninkov87},$^5$\citet{munos08},$^6$\citet{hynes04},$^7$\citet{o02},$^8$\citet{park04},$^9$\citet{oro02},$^{10}$\citet{san99},$^{11}$\citet{sob99},$^{12}$\citet{oro04},$^{13}$\citet{homan06},$^{14}$\citet{mcc07},$^{15}$\citet{fil01},$^{16}$\citet{zur02}}
\label{fintab}
\end{deluxetable}

\newpage

\begin{figure}[ptbptbptb]
\includegraphics[scale=1.0,angle=-90]{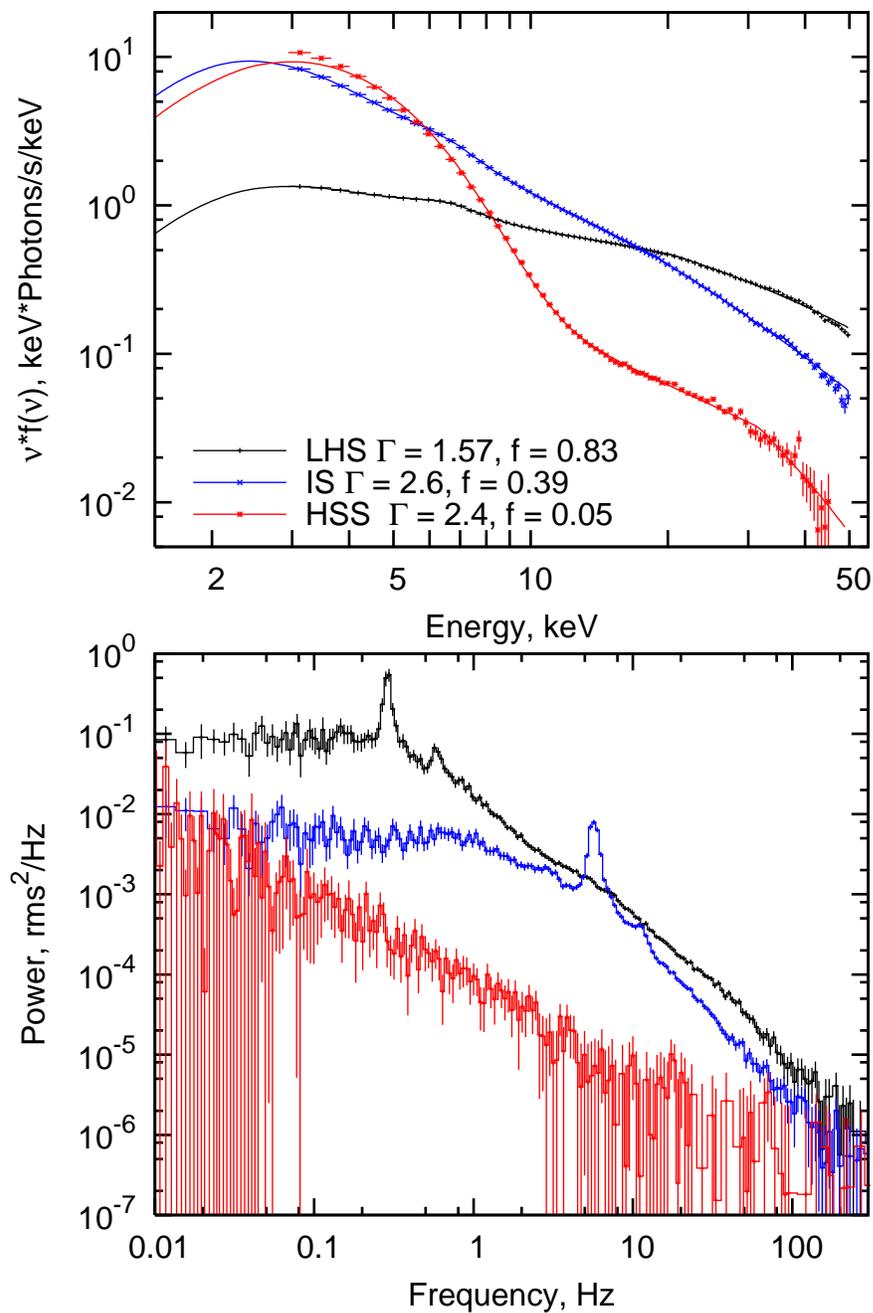}
\caption{Three representative energy ({\it top}) and power ({\it bottom}) spectra during the rise part of the 1998 outburst of XTE J1550-564. Data are taken from {\it RXTE} observations 30188-06-01-00 ({\it red}), 30188-06-01-00 ({\it blue}) and 30191-01-05-00 ({\it black}). 
In the top panel the solid curves correspond to the best-fit model spectra.
}
\label{1550_evolution}
\end{figure}

\clearpage

\begin{figure}[ptbptbptb]
\includegraphics[scale=0.65,angle=-90]{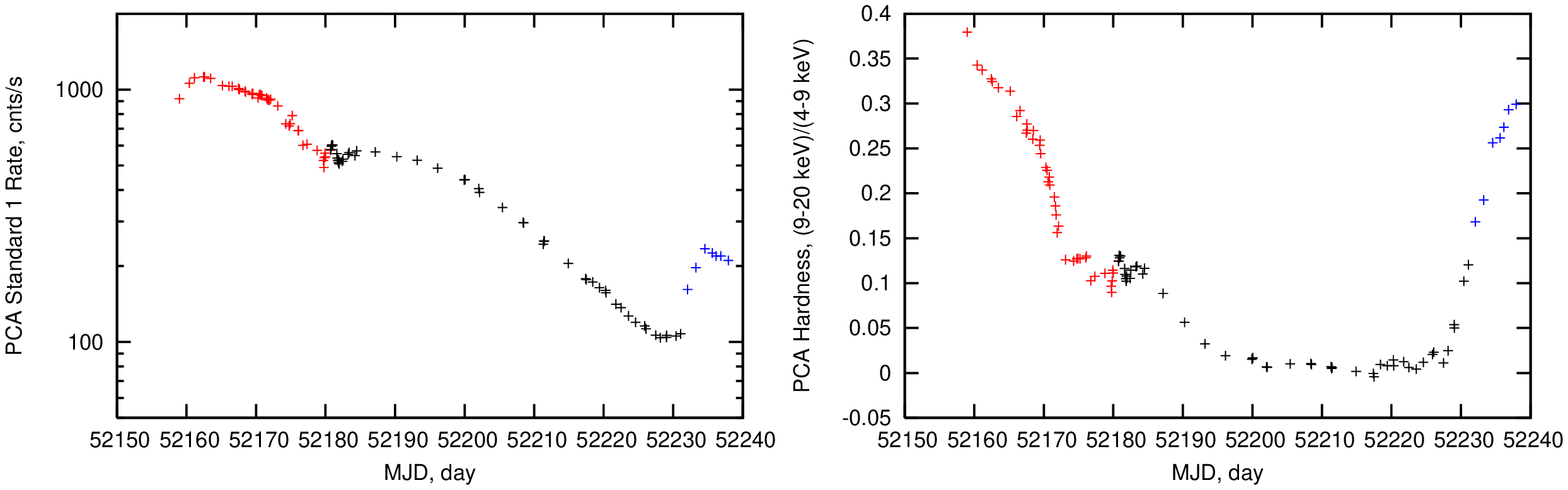}
\includegraphics[scale=0.65,angle=-90]{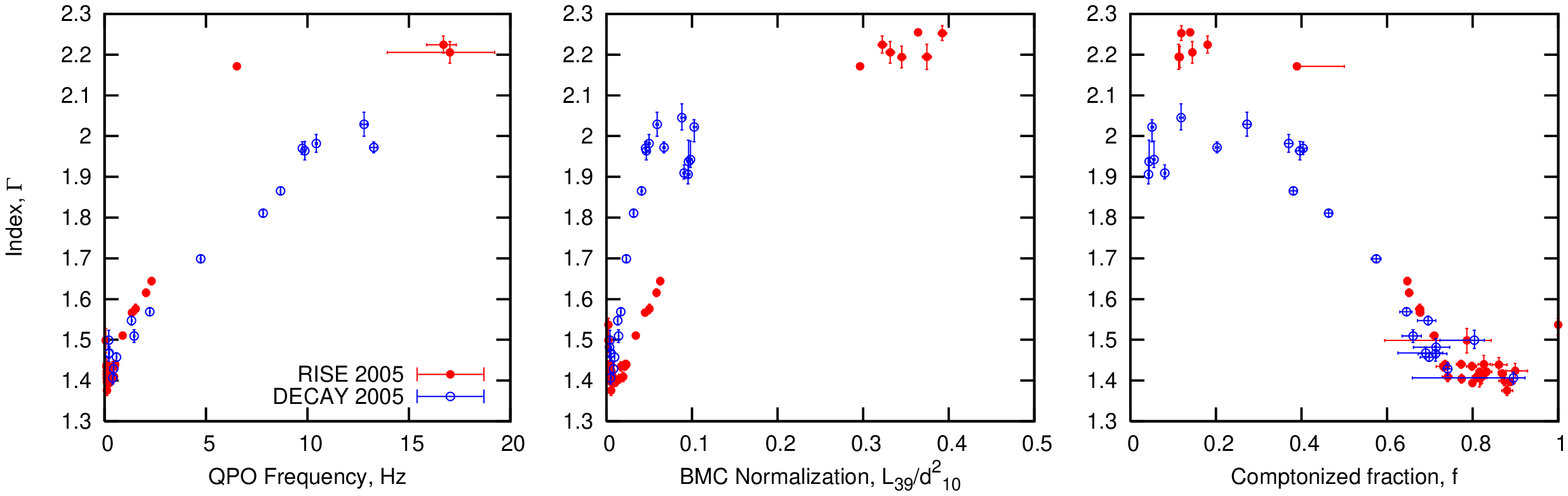}
\caption{{\it Top:} PCA Standard1 count rate (left) and hardness (right) for the
2005 outbursts from GRO J1655-40. Blue  and red colors stand for rise and decay transitions  correspondingly.
{\it Botom}: Photon index is plotted versus QPO frequency (left),
BMC normalization (middle) and Comptonized fraction (right) 
for the rise (red) and the decay (blue) transitions. }
\label{1655_lc}
\end{figure}

\clearpage

\begin{figure}[ptbptbptb]
\includegraphics[scale=0.65,angle=-90]{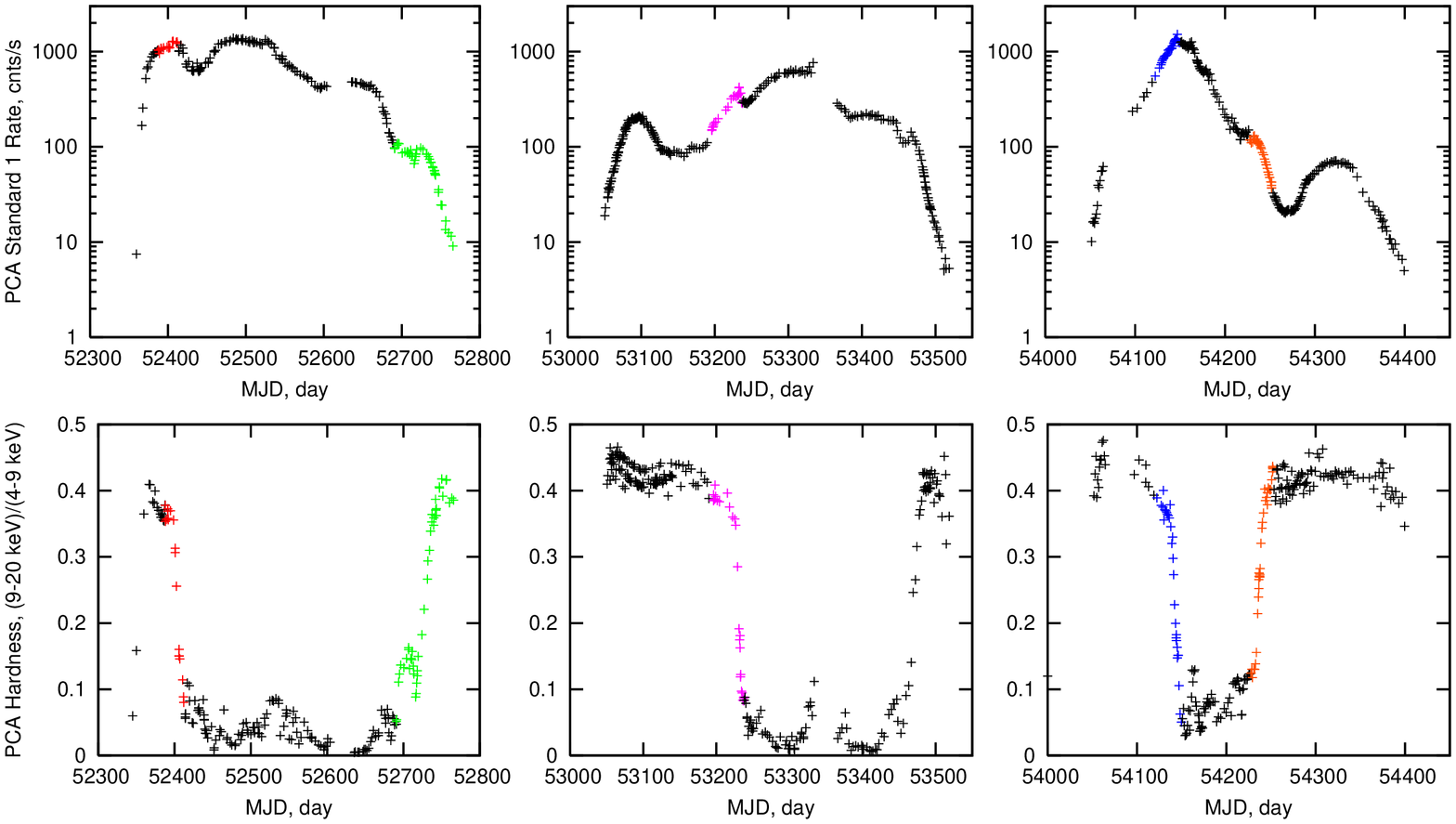}
\includegraphics[scale=0.65,angle=-90]{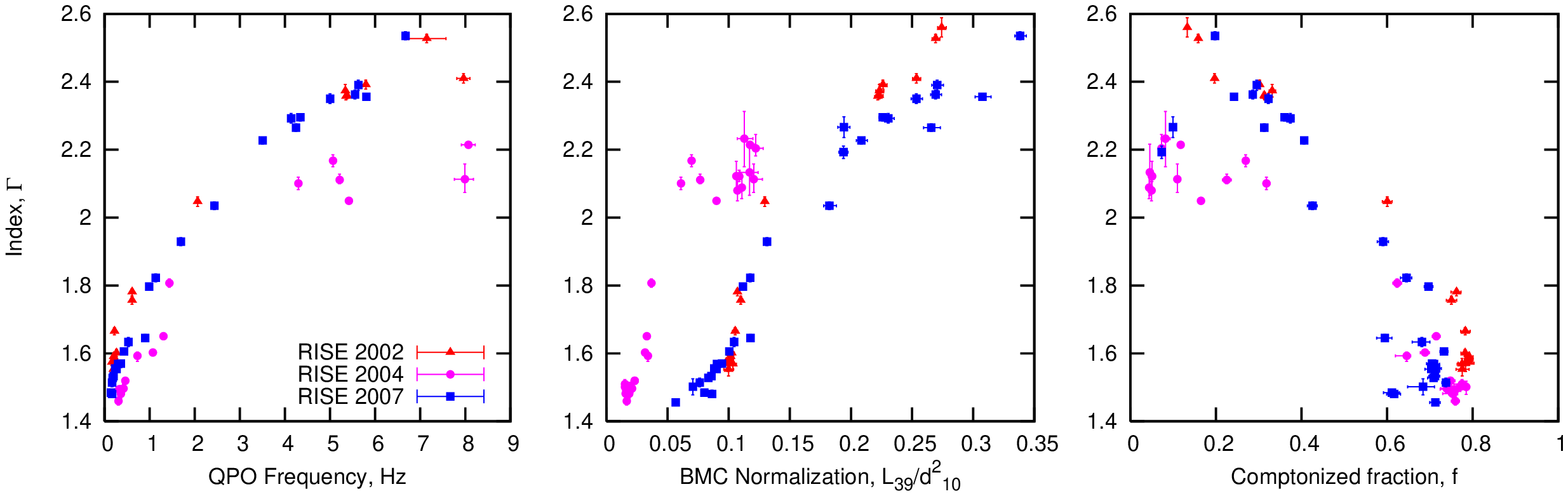}
\includegraphics[scale=0.65,angle=-90]{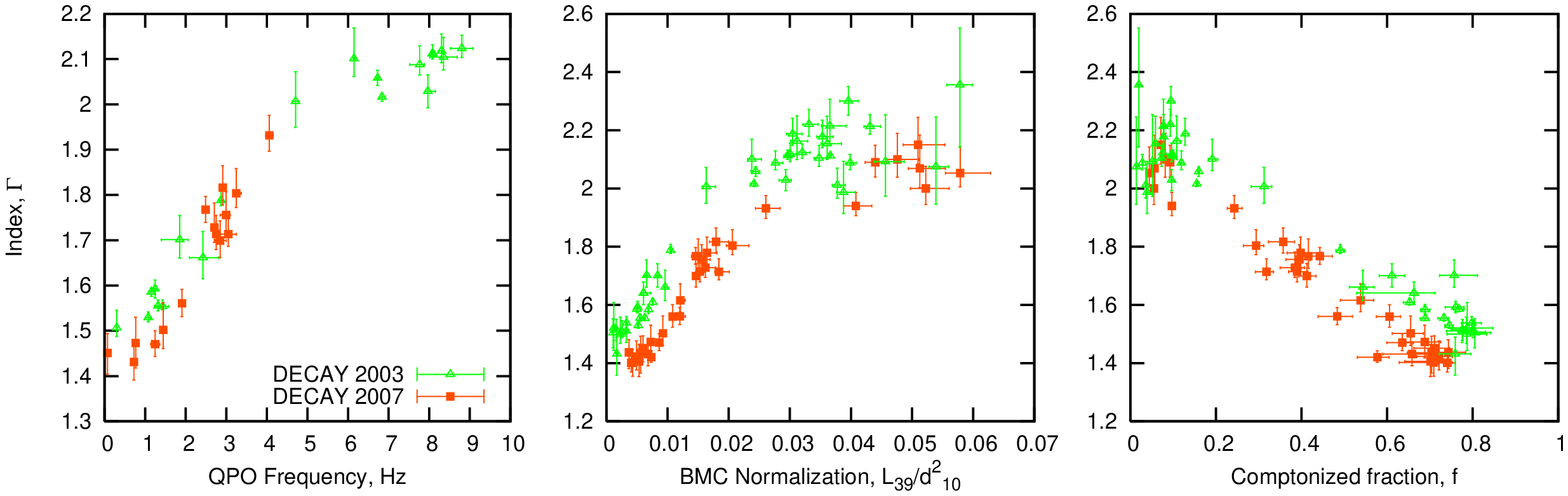}
\caption{ PCA Standard1 count rate ({\it top row}) and hardness ({\it second row})
 for three outbursts from GX 339-4 on 2002 (left), 2004 (middle) and 2007 (right).
{\it Third and bottom rows:} Photon index plotted versus QPO frequency (left),
BMC normalization (middle) and Comptonized fraction (right) 
for transitions in GX 339-4 (each transition is indicated by different color).}
\label{gx339_lc}
\end{figure}





\clearpage

\begin{figure}[ptbptbptb]
\includegraphics[scale=0.65,angle=-90]{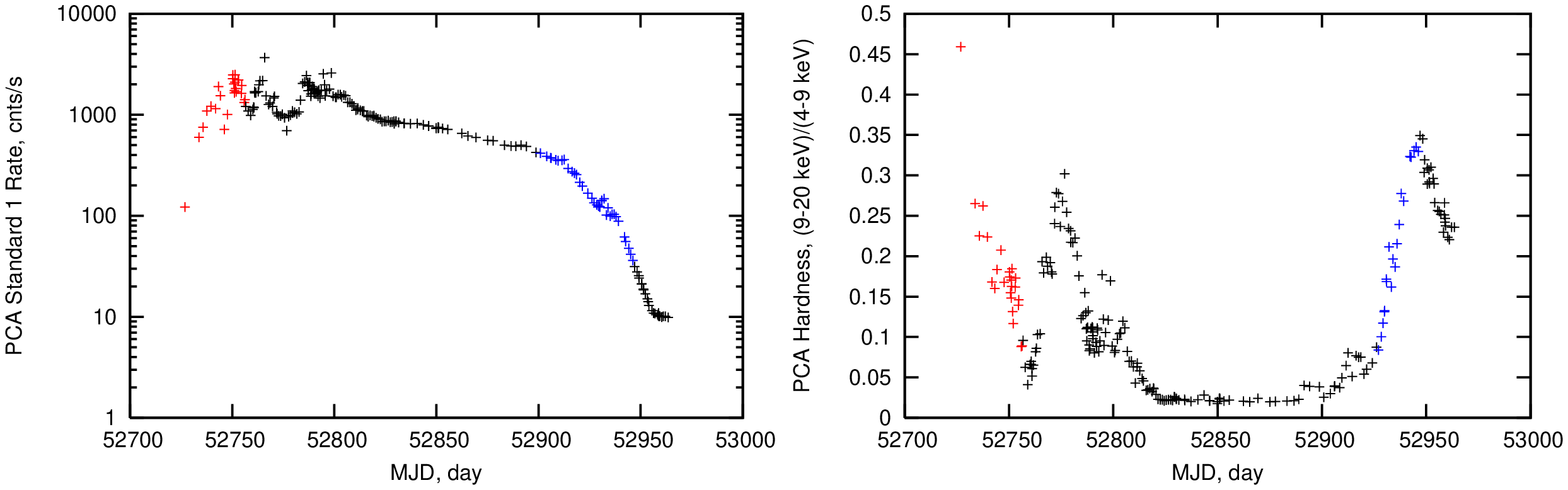}
\includegraphics[scale=0.65,angle=-90]{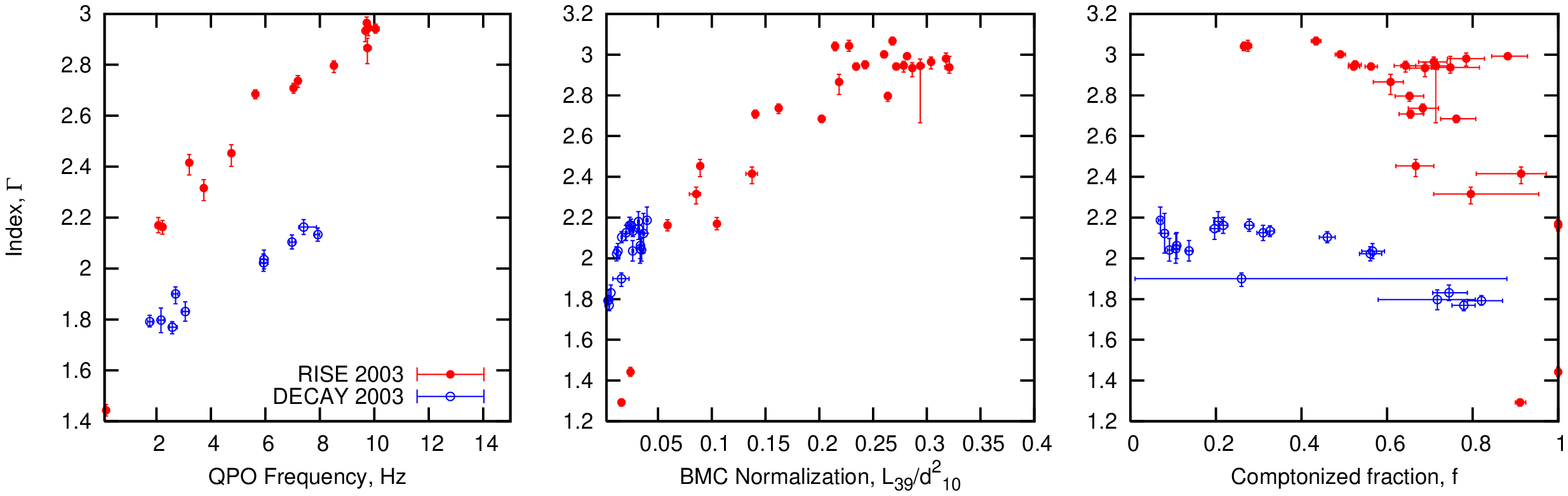}
\caption{{\it Top}: PCA Standard1 count rate (left) and hardness (right) for the
2003 outbursts from H1743-322. {\it Bottom}
Photon index plotted versus QPO frequency (left),
BMC normalization (middle) and Comptonized fraction (right) 
for the rise and the decay transitions.}
\label{1743_lc}
\end{figure}



\clearpage

\begin{figure}[ptbptbptb]
\includegraphics[scale=0.65,angle=-90]{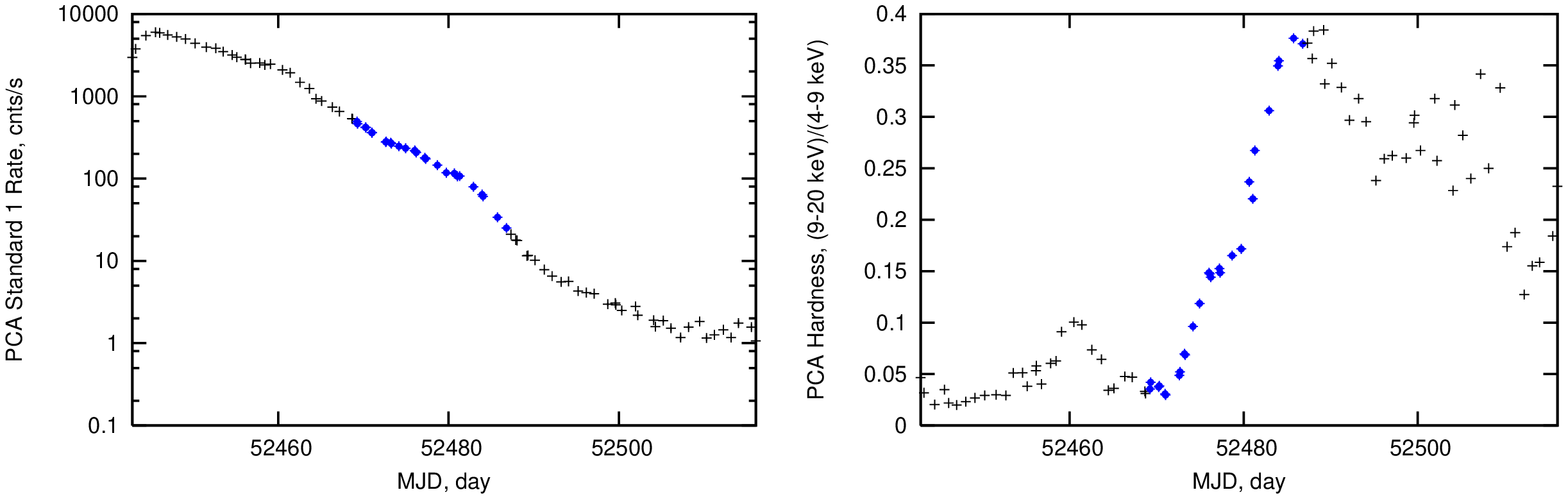}
\includegraphics[scale=0.65,angle=-90]{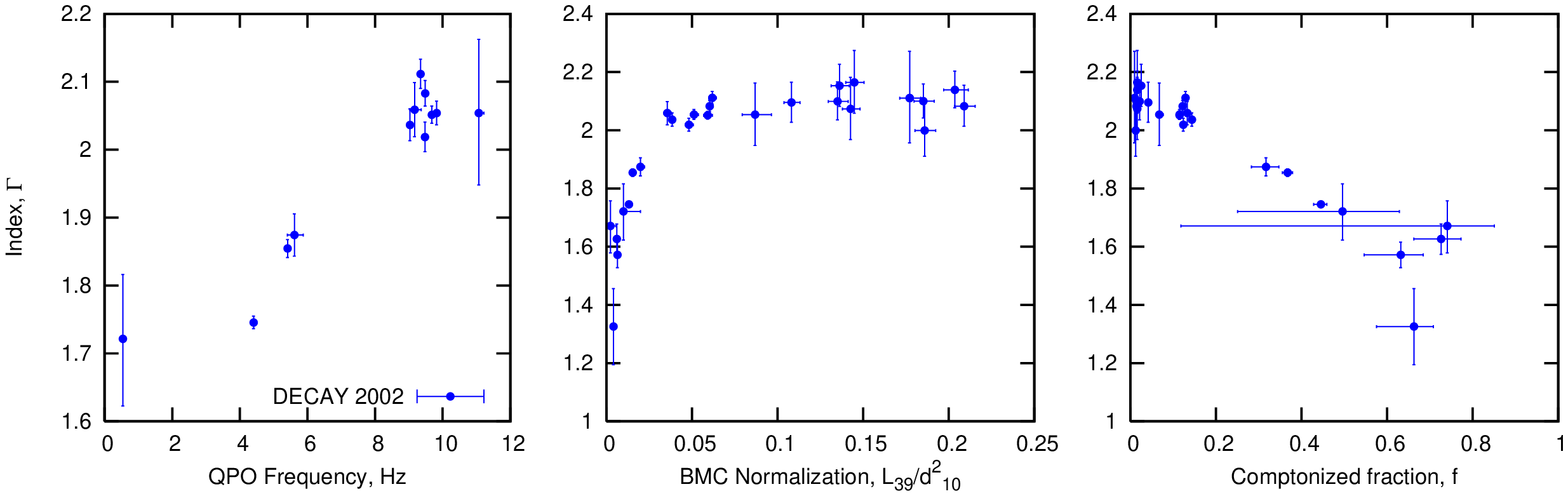}
\caption{{\it Top}: PCA Standard1 count rate (left) and hardness (right) for the
2002 outburst from 4U 1543-47. {\it Bottom}: Photon  index versus QPO (left), 
BMC normalization (middle) and Comptonized fraction (right) 
for the decay transition.}
\label{1543_lc}
\end{figure}

\clearpage

\begin{figure}[ptbptbptb]
\includegraphics[scale=0.65,angle=-90]{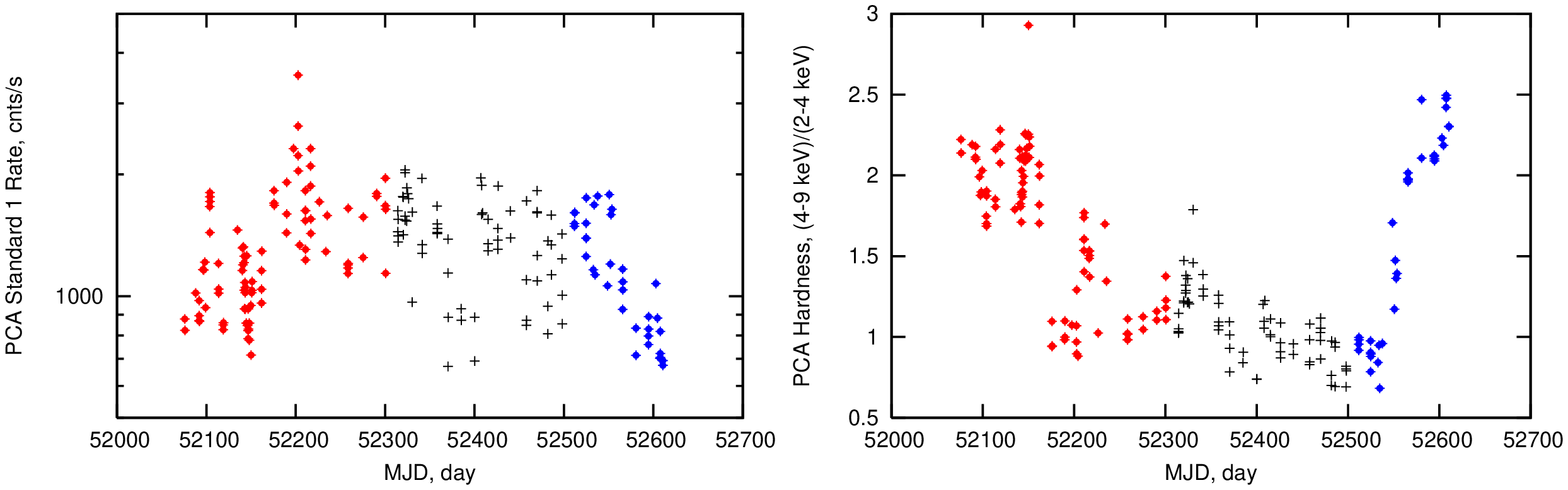}
\includegraphics[scale=0.65,angle=-90]{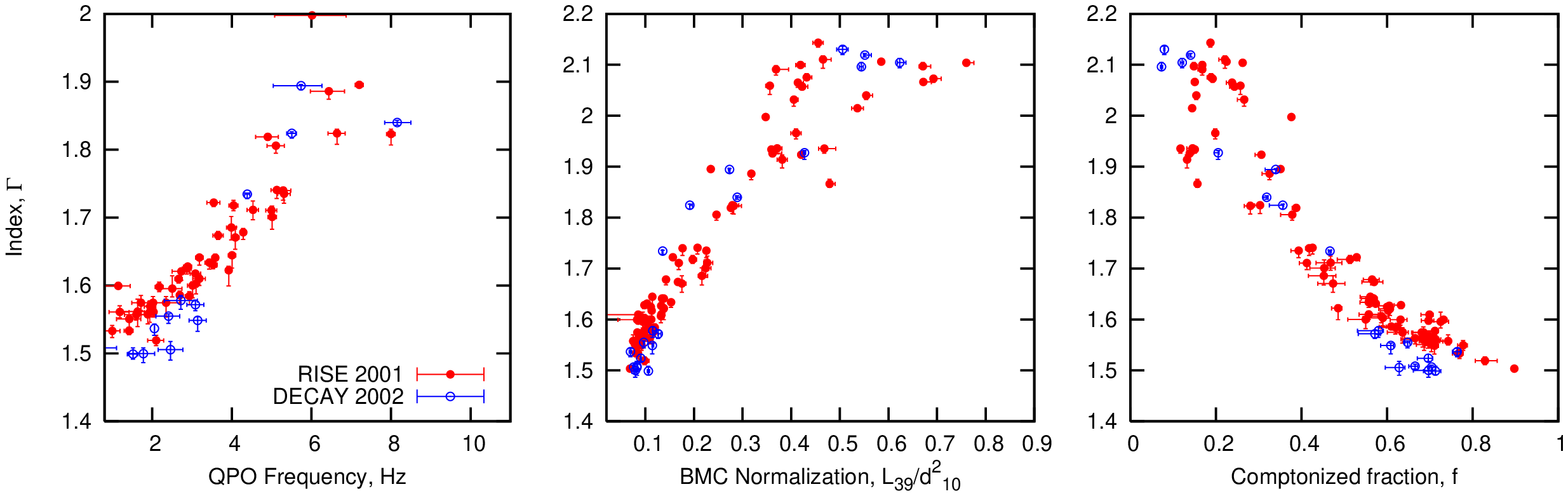}
\caption{{\it Top}: PCA Standard1 lighcurve (left) and hardness (right) for the
Cygnus X-1 during 2001-2002. Red points indicate soft-to-hard transition
while blue points show hard-to-soft transition. Observations during
the HS are shown in black color.{\it Bottom}:
Photon index versus QPO (left),
BMC normalization (middle) and Comptonized fraction (right) 
during 2001-2002 state transitions in Cygnus X-1.}
\label{cygx1_lc}
\end{figure}



\clearpage

\begin{figure}[ptbptbptb]
\includegraphics[scale=0.5,angle=-90]{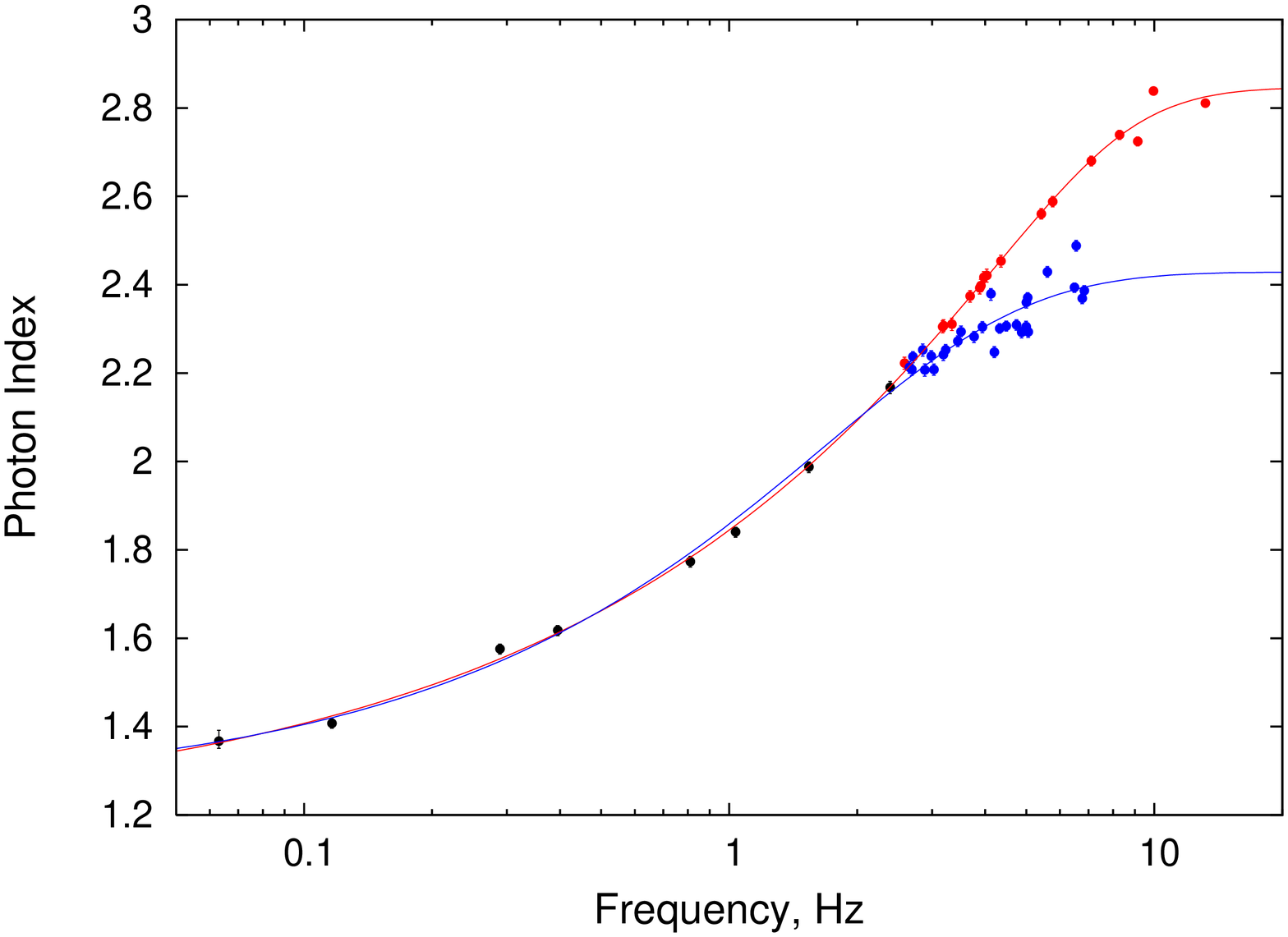}
\includegraphics[scale=0.5,angle=-90]{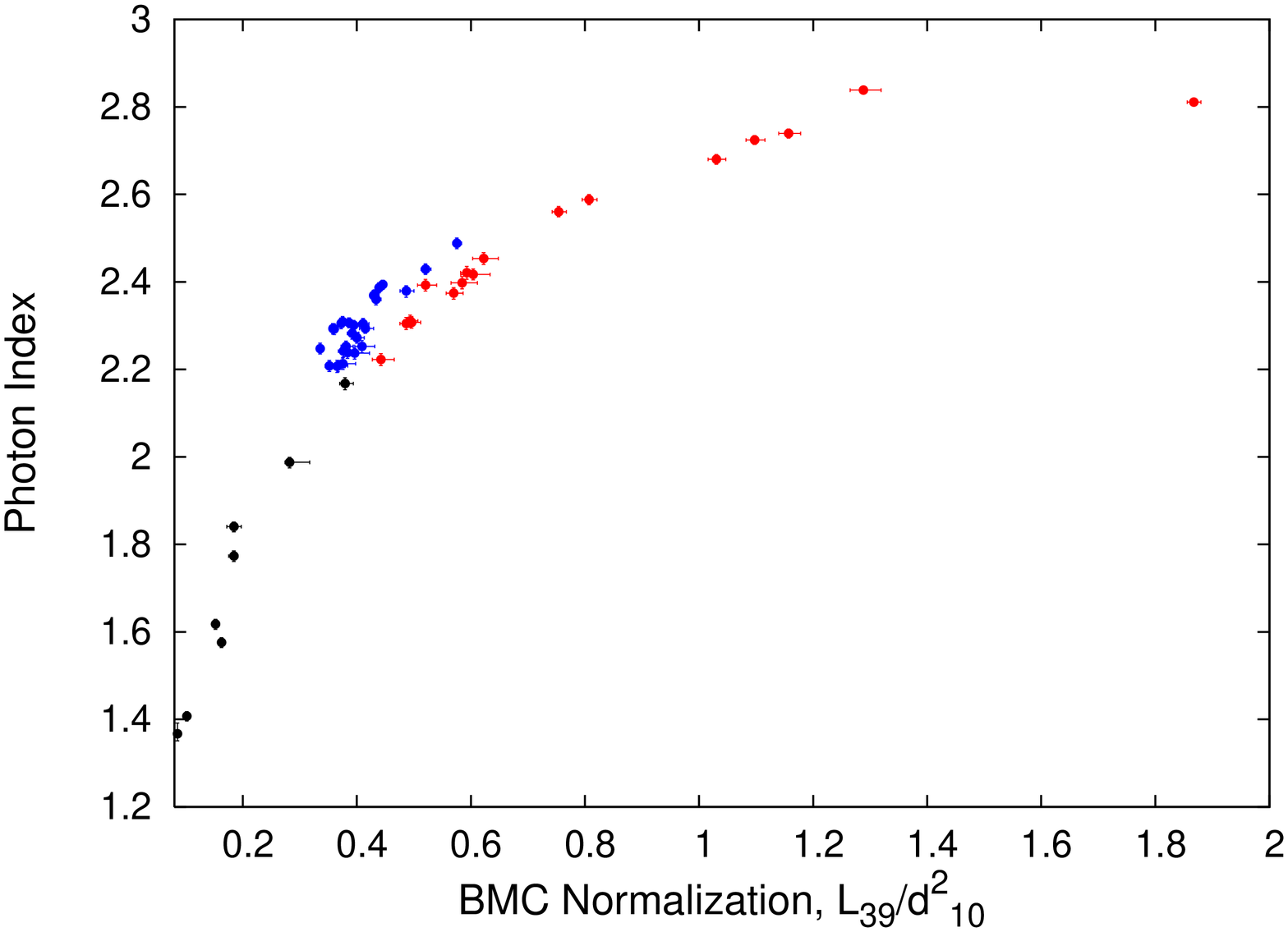}
\caption{Photon Index - QPO frequency and Index-BMC normalization correlations (upper and bottom panels respectively) during the rise 
of XTE J1550-564 1998 outburst. Index-QPO correlation exhibits two different 
tracks, which correspond to different index saturation levels (see text for details).}
\label{1550_tracks}
\end{figure}


\clearpage

\begin{figure}[ptbptbptb]
\includegraphics[scale=0.425,angle=-90]{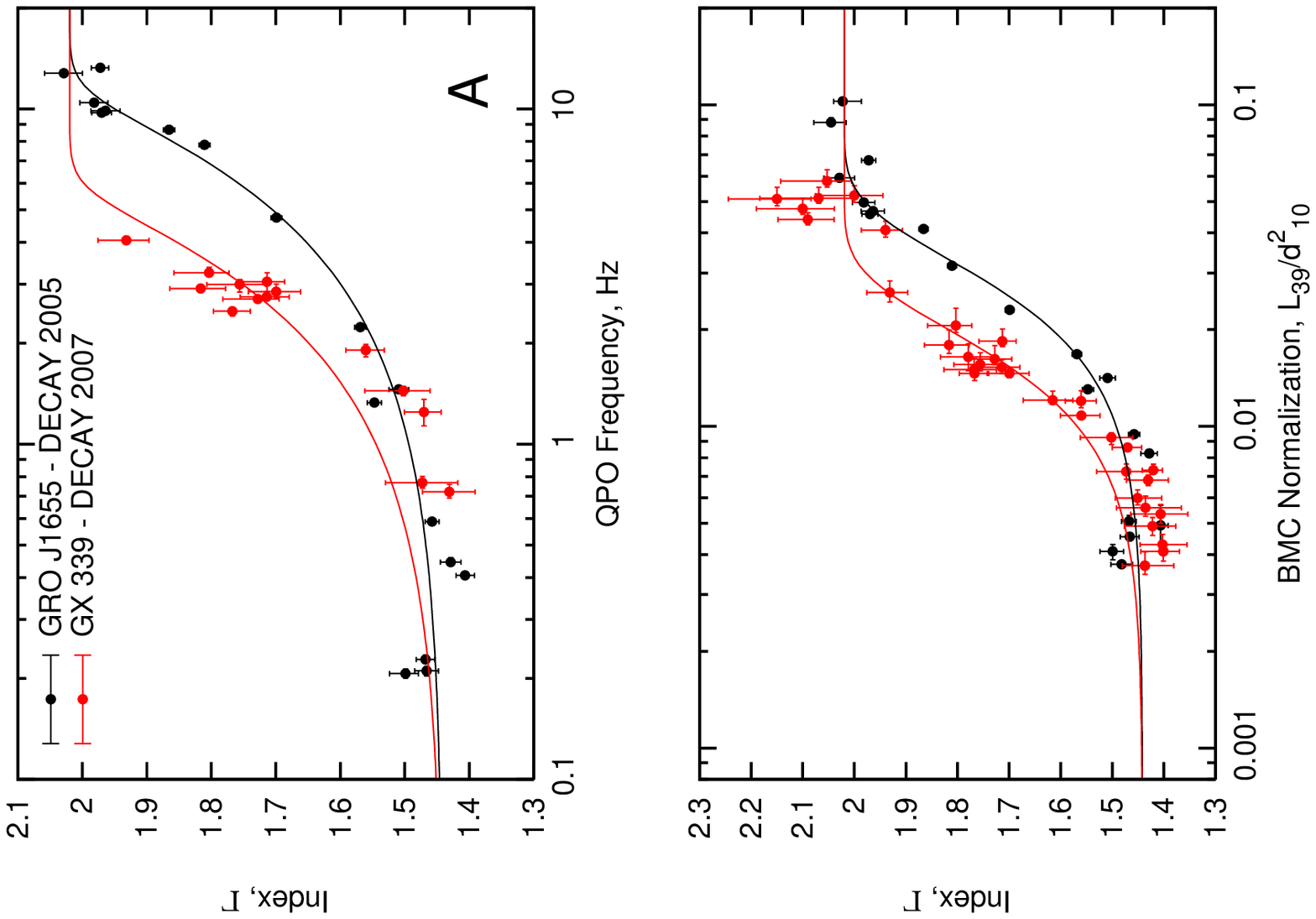}
\includegraphics[scale=0.425,angle=-90]{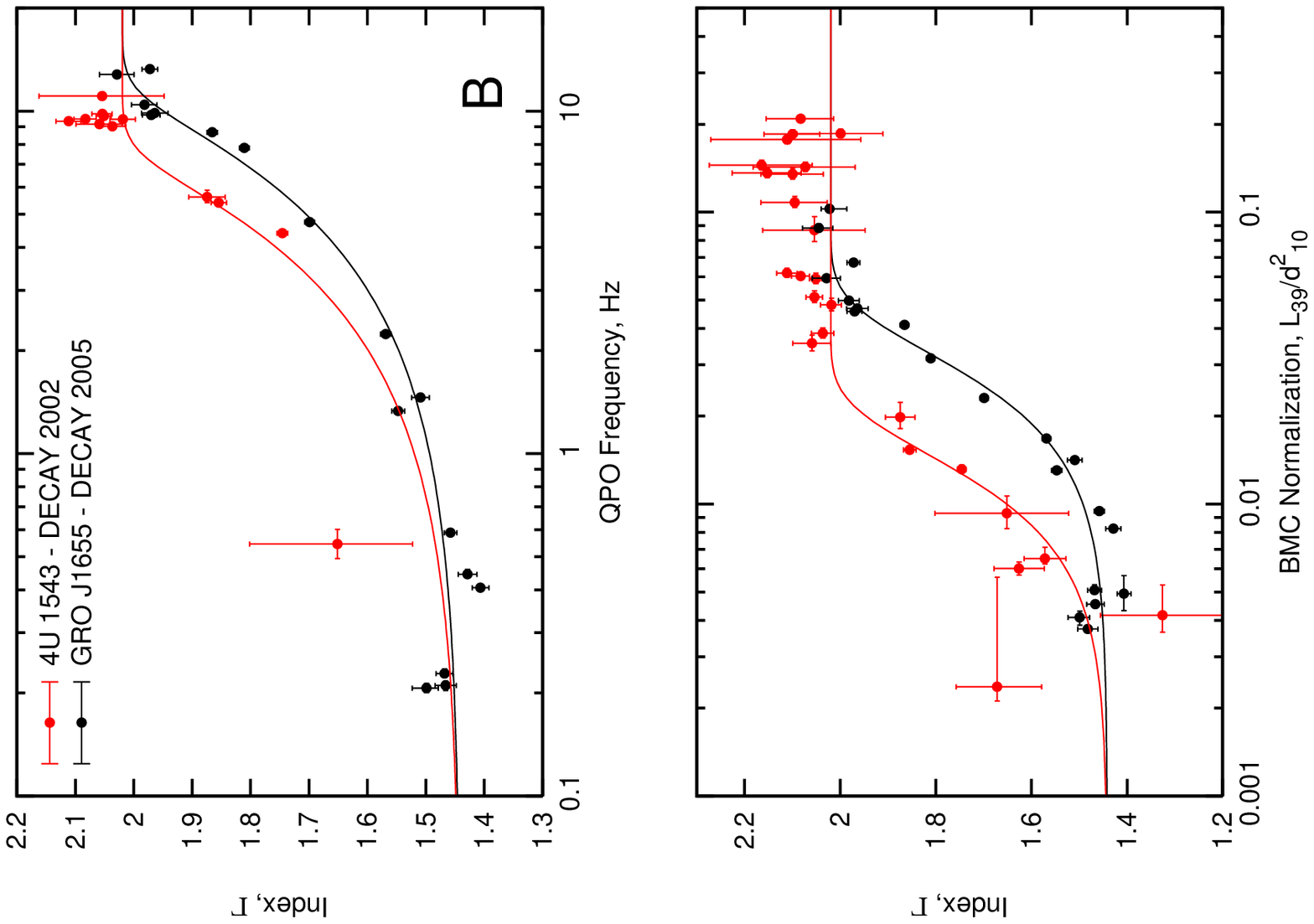}
\includegraphics[scale=0.425,angle=-90]{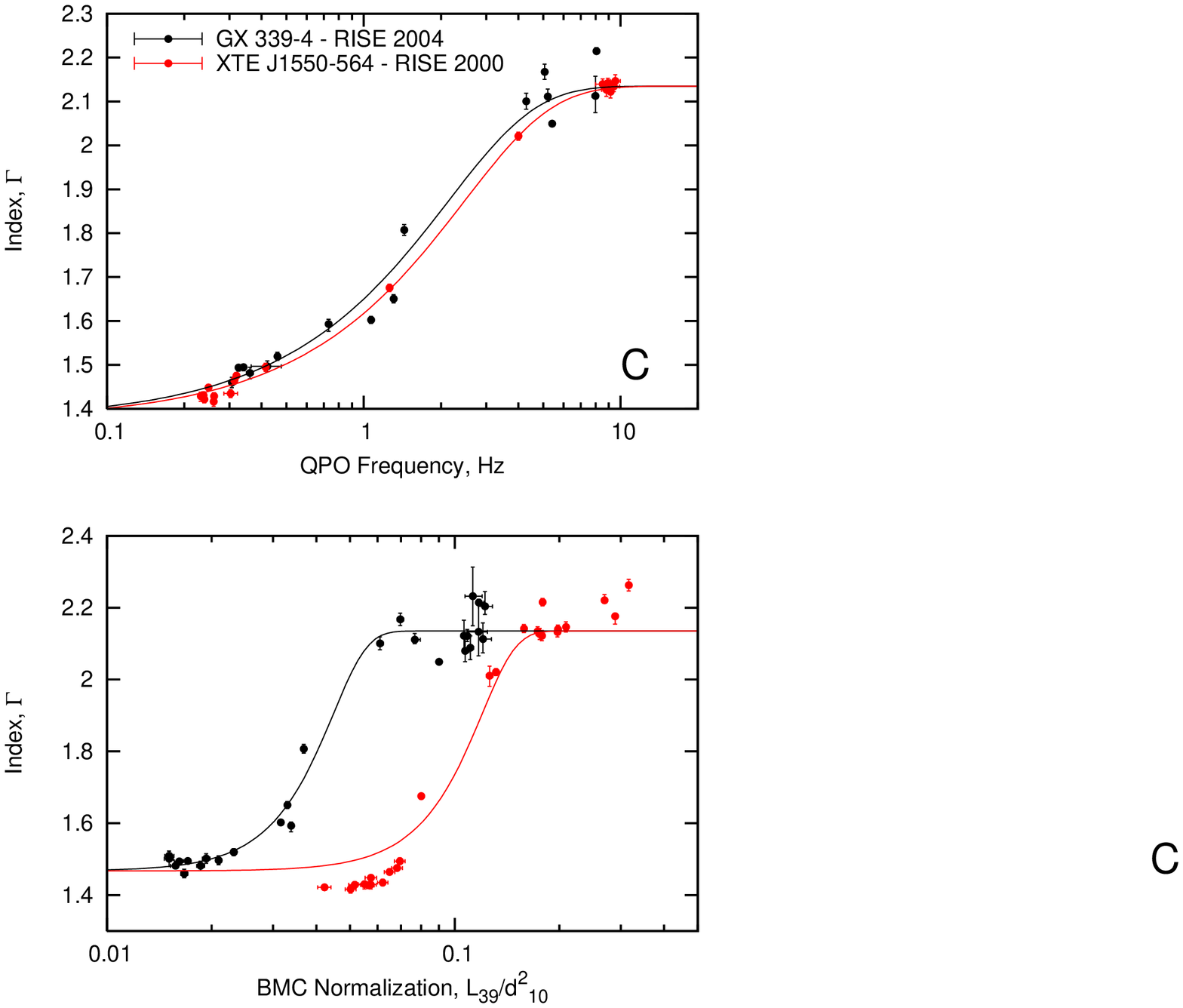}
\vspace{0.1in}

\includegraphics[scale=0.425,angle=-90]{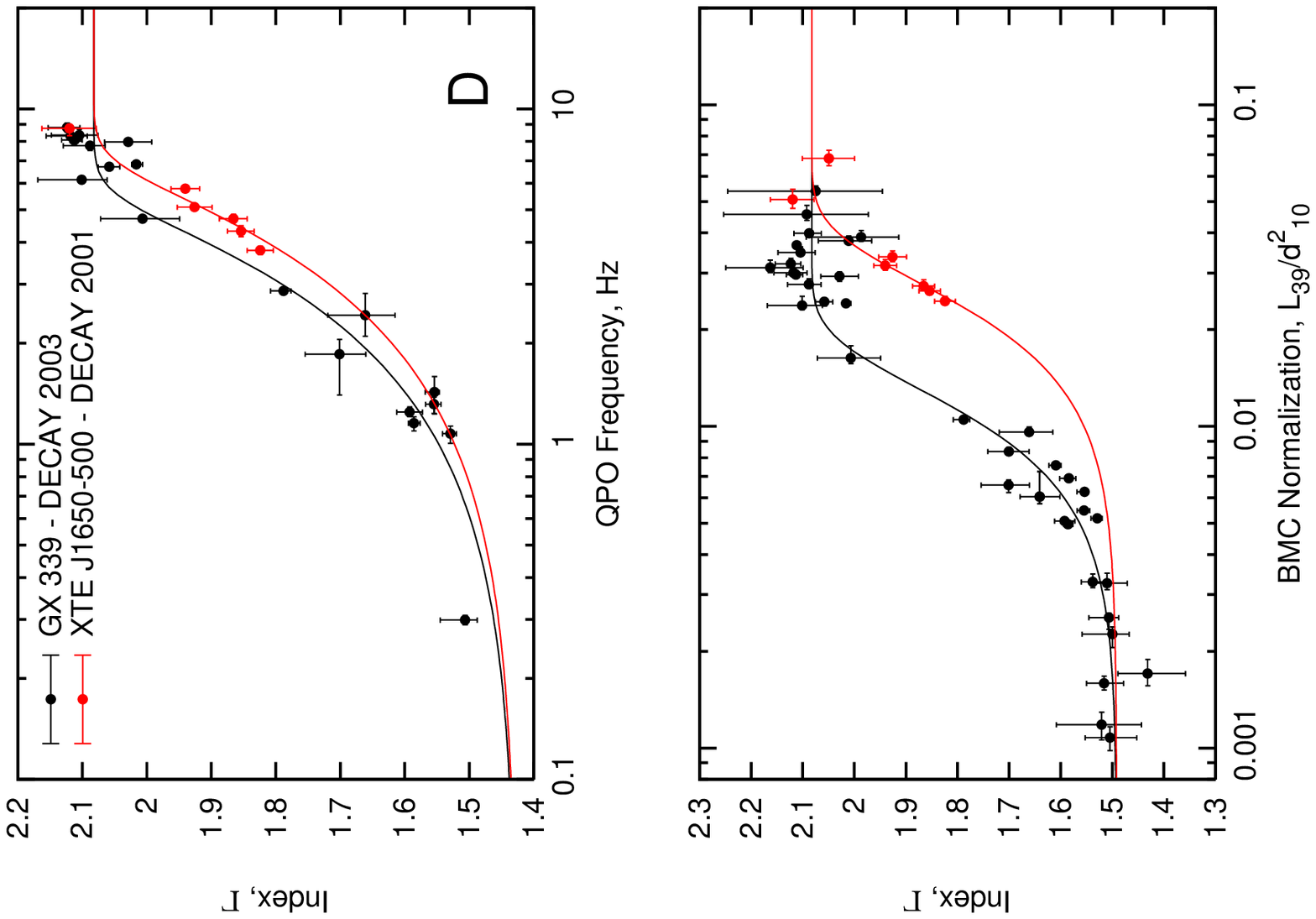}
\includegraphics[scale=0.425,angle=-90]{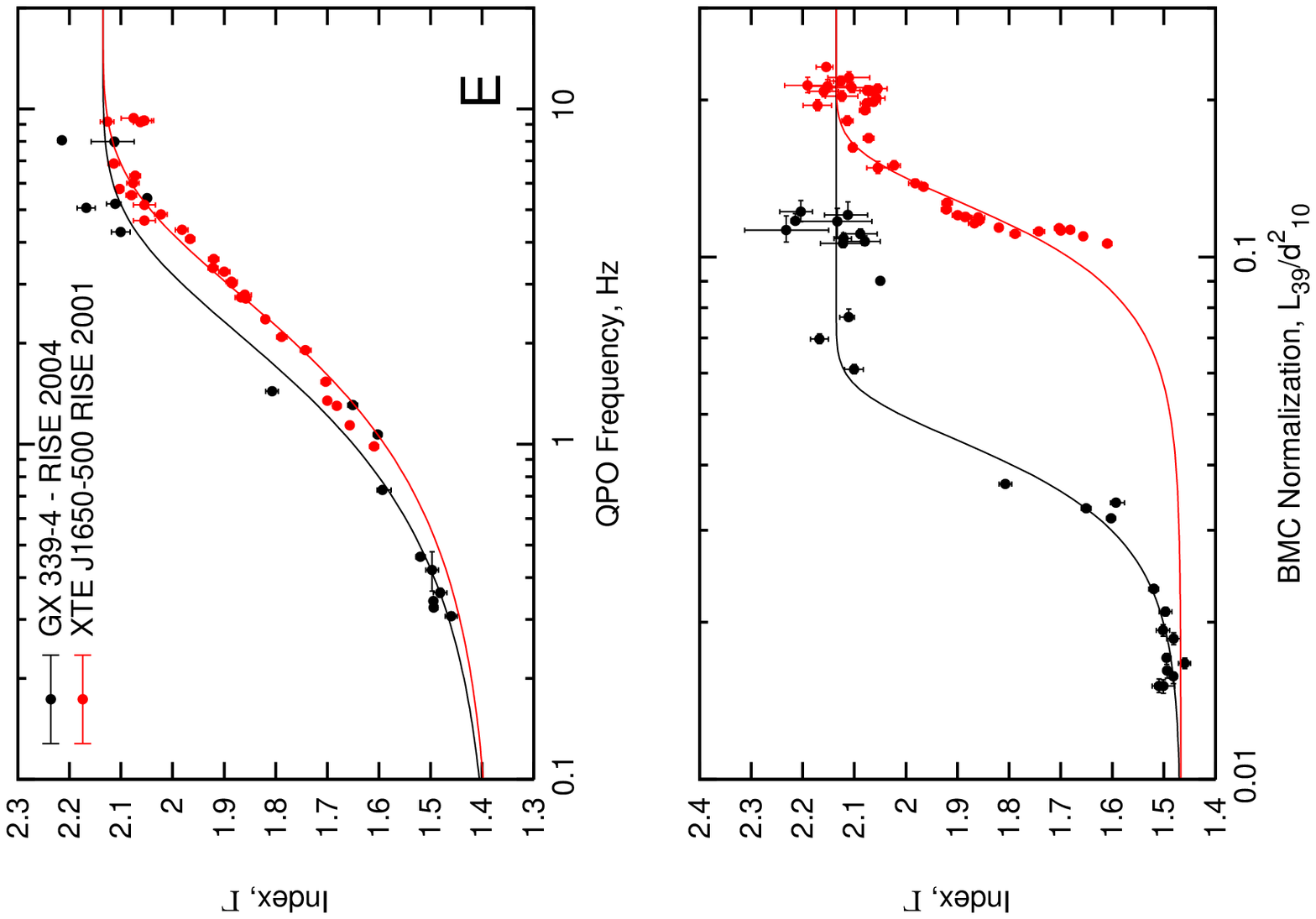}
\includegraphics[scale=0.425,angle=-90]{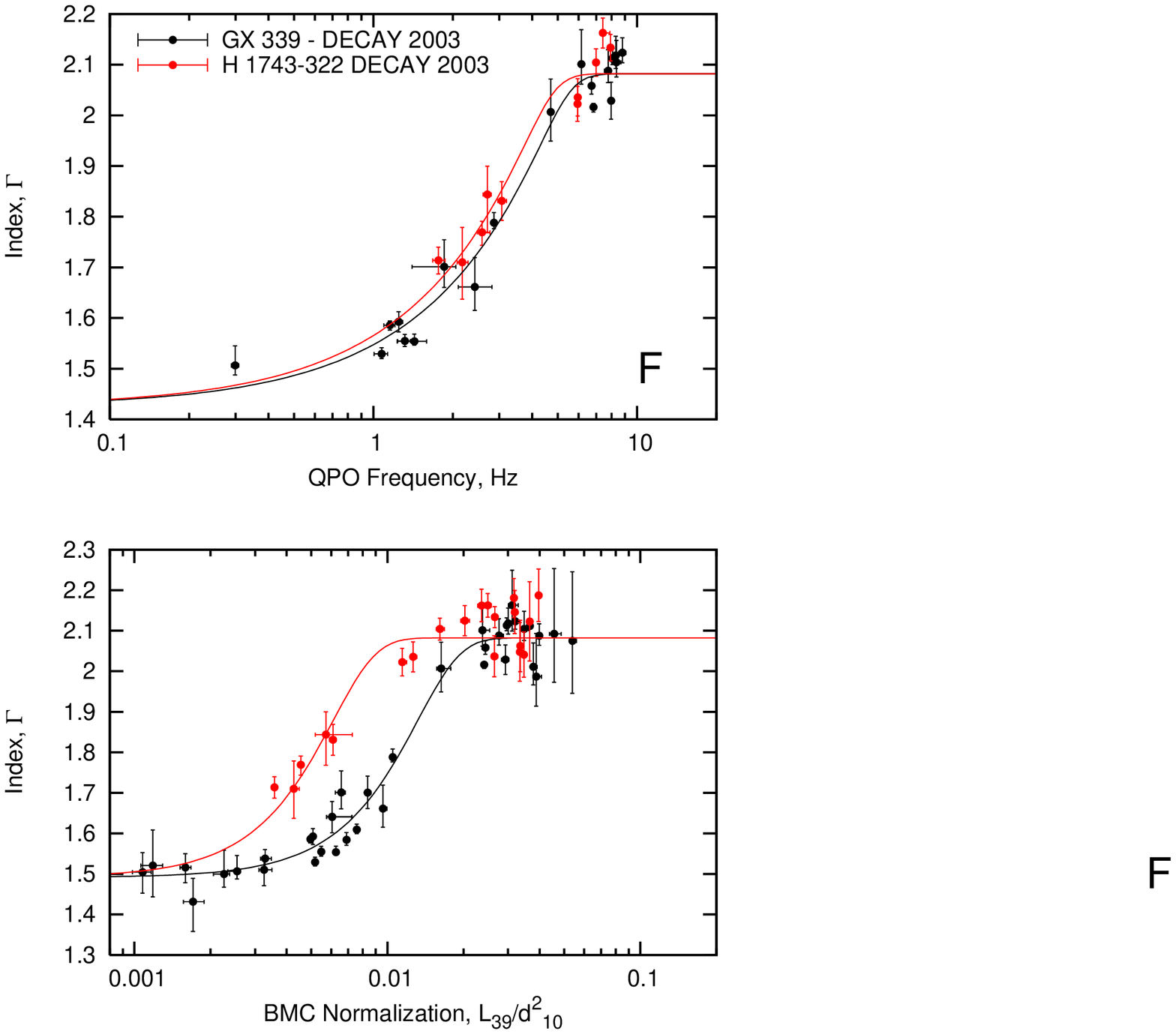}
\caption{Scaling photon index  vs QPO frequency  and normalization  correlations
 for  BH sources. Normalization plot is placed below the corresponding
QPO frequency panel. In these diagrams the target transition for scaling
is shown in red color while reference transition is plotted in black.
Each pair (from left to right and from top to bottom) corresponds to the 
row in Table \ref{scaling_tab} beginning with the second row.}
\label{shift_figs}
\end{figure}

\clearpage

\begin{figure}[ptbptbptb]
\includegraphics[scale=0.425,angle=-90]{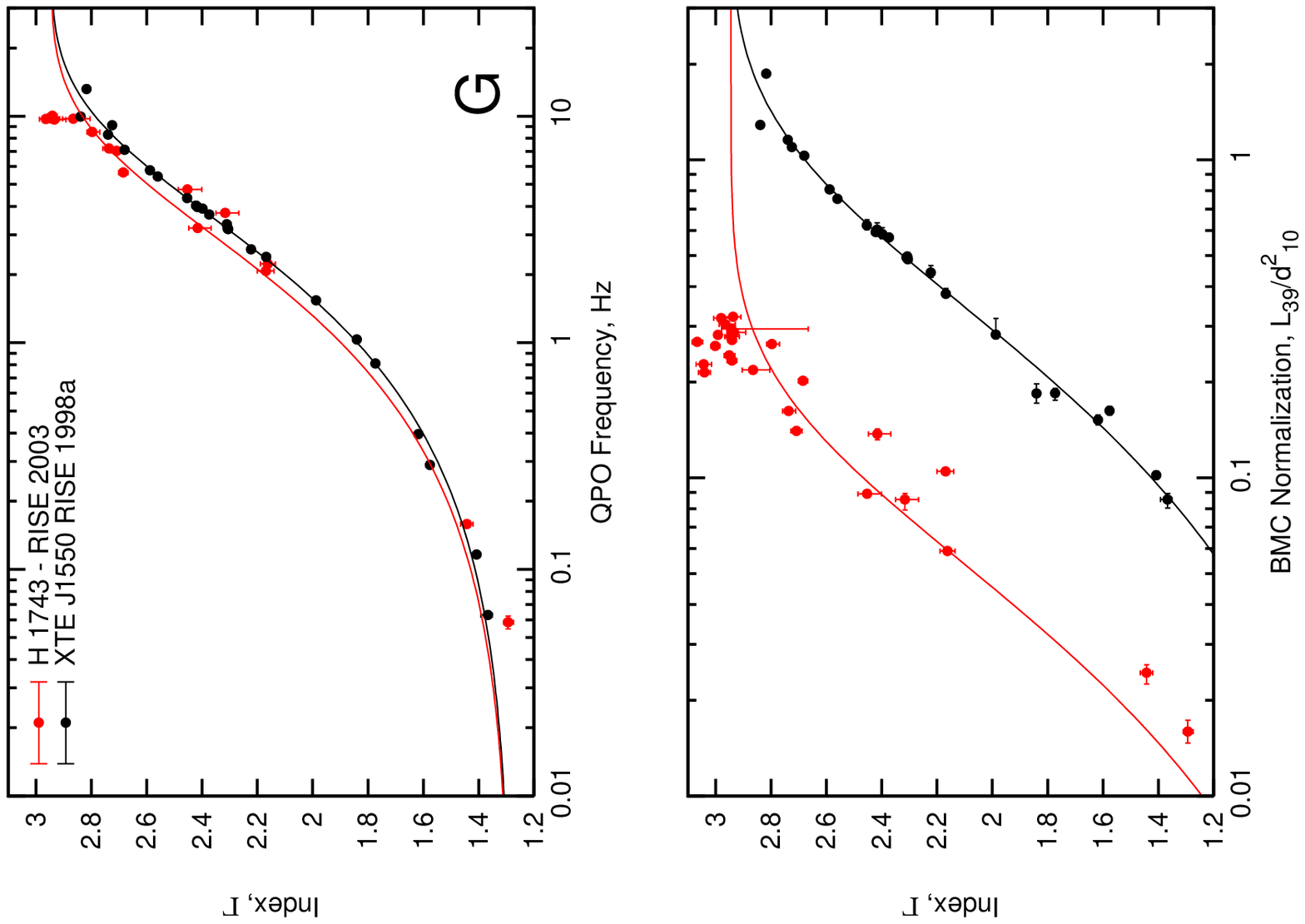}
\includegraphics[scale=0.425,angle=-90]{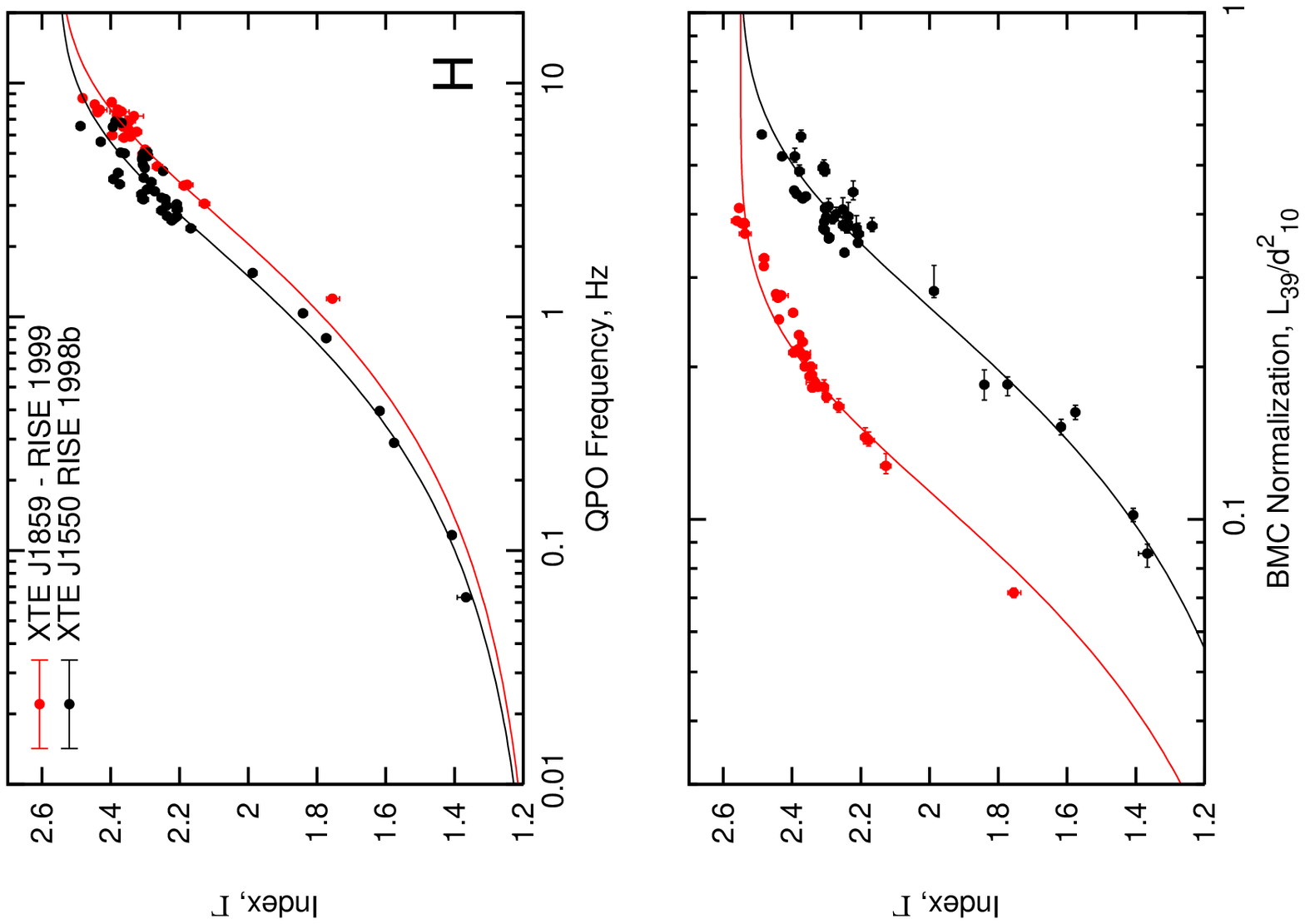}
\includegraphics[scale=0.425,angle=-90]{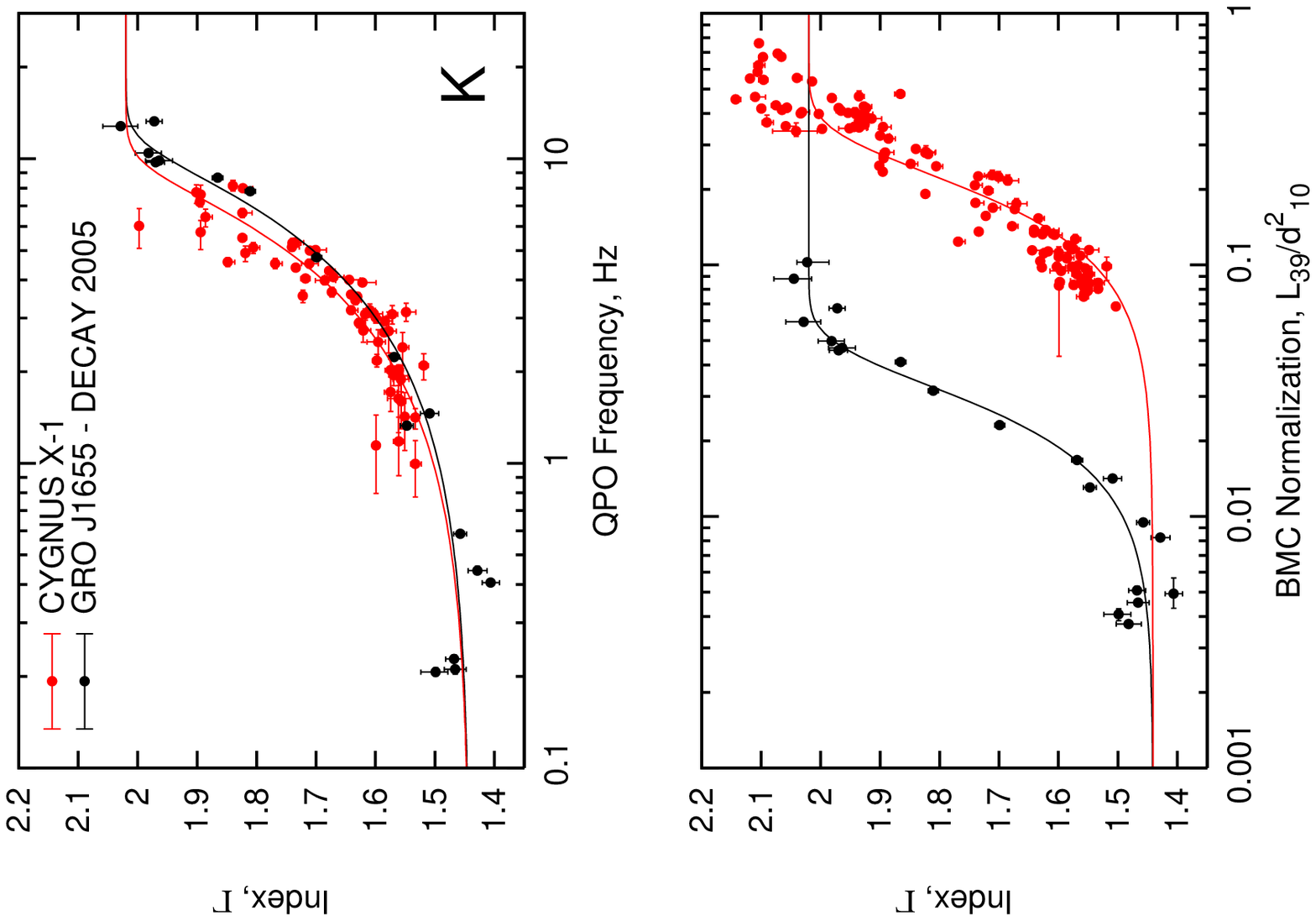}
\caption{Index-QPO frequency and index-normalization scaling diagrams
continued.}
\label{shift_figs_2}
\end{figure}


\begin{thebibliography}{}


\bibitem[Belloni et al.(2005)]{bel05} Belloni, T., Homan, J., 
Casella, P., van der Klis, M., Nespoli, E., Lewin, W.~H.~G., Miller, J.~M., 
\& M{\'e}ndez, M.\ 2005, \aap, 440, 207 

\bibitem[Belloni(2005)]{bell05} Belloni, T.\ 2005, Interacting 
Binaries: Accretion, Evolution, and Outcomes, 797, 197 

\bibitem[Belloni et al.(2006)]{bel06} Belloni, T., et al.\ 
2006, \mnras, 367, 1113 

\bibitem[Borozdin \etal (1999)]{b99}
Borozdin, K., Revnivtsev, M.,  Trudolyubov, S., Shrader, C. \& Titarchuk, L. 1999, \apj, 517, 367 


\bibitem[Casella et al.(2004)]{cas04} Casella, P., Belloni, 
T., Homan, J., \& Stella, L.\ 2004, \aap, 426, 587 


\bibitem[Dickey \& Lockman (1990)]{nH} Dickey,J.M., \& Lockman, F.J. 1990, ARAA, 28, 215


\bibitem[Dewangan, Titarchuk \&  Griffiths (2006)]{dtg06} Dewangan, G. C., Titarchuk, L., Griffiths, R. E. 2006, \apj, 637, 1, L21

\bibitem[Filippenko \& Chornock(2001)]{fil01} Filippenko, 
A.~V., \& Chornock, R.\ 2001, \iaucirc, 7644, 2 

\bibitem[Fiorito \& Titarchuk (2004)]{ft04} Fiorito, R., Titarchuk, L. 2004, \apj, 614,  L113



\bibitem[Green, Bailyn \& Orosz (2001)]{greene01} Greene, J., Bailyn, C. D., \& Orosz, J. A. 2001, \apj, 554, 1290


\bibitem[Herrero \etal (1995)]{her95} Herrero, J., \etal 1995, \aap, 297, 556

\bibitem[Hjellming \& Rupen (1995)]{hr95} Hjellming, R.M.,\& Rupen, M.P. 1995, \nat, 375,464 


\bibitem[Homan et al.(2006)]{homan06} Homan, J., Wijnands, R., 
Kong, A., Miller, J.~M., Rossi, S., Belloni, T., 
\& Lewin, W.~H.~G.\ 2006, \mnras, 366, 235 

	

\bibitem[Hynes et al.(2004)]{hynes04} Hynes, R.~I., Steeghs, 
D., Casares, J., Charles, P.~A., \& O'Brien, K.\ 2004, \apj, 609, 317 





\bibitem[Kalemci et al.(2005)]{kal05} Kalemci, E., Tomsick, 
J.~A., Buxton, M.~M., Rothschild, R.~E., Pottschmidt, K., Corbel, S., 
Brocksopp, C., \& Kaaret, P.\ 2005, \apj, 622, 508 

\bibitem[Klein-Wolt  \& van der Klis(2008)]{kw08} Klein-Wolt, M., \& van der Klis, M.\ 2008, \apj, 675, 1407 



\bibitem[Lang (1998)]{lang98} Lang, K. R., 1998, {Astrophysical Formulae},
(Berlin: Springer)

\bibitem[Laurent \& Titarchuk (1999)]{lt99}
Laurent, P. \& Titarchuk, L.   1999,  \apj, 511, 289  (LT99)

\bibitem[Laurent \& Titarchuk (2007)]{lt07}
Laurent, P. \& Titarchuk, L.   2007,  \apj, 656, 1056 

\bibitem[McClintock et al.(2007)]{mcc07} McClintock, J.~E., 
Remillard, R.~A., Rupen, M.~P., Torres, M.~A.~P., Steeghs, D., Levine, 
A.~M., \& Orosz, J.~A.\ 2007, ArXiv e-prints, 705, arXiv:0705.1034 

\bibitem[Montanari et al.  (2008)]{mtf08}
Montanari, E.,  Titarchuk, L. \& Frontera, F. 2008,  ApJ, accepted, (arXiv:0810.5720) 
 

\bibitem[Mu{\~n}oz-Darias et al.(2008)]{munos08} 
Mu{\~n}oz-Darias, T., Casares, J., 
\& Mart{\'{\i}}nez-Pais, I.~G.\ 2008, \mnras, 385, 2205 




\bibitem[Ninkov et al.(1987)]{ninkov87} Ninkov, Z., Walker, 
G.~A.~H., \& Yang, S.\ 1987, \apj, 321, 425 


\bibitem[Orosz(2003)]{o02} Orosz, J.~A.\ 2003, A Massive 
Star Odyssey: From Main Sequence to Supernova, 212, 365 (astro-ph/0209041)



\bibitem[Orosz et al.(2002)]{oro02} Orosz, J.~A., et al.\ 
2002, \apj, 568, 845 

\bibitem[Orosz et al.(2004)]{oro04} Orosz, J.~A., McClintock, 
J.~E., Remillard, R.~A., \& Corbel, S.\ 2004, \apj, 616, 376 

\bibitem[Park et al.(2004)]{park04} Park, S.~Q., et al.\ 2004, 
\apj, 610, 378 

\bibitem[Remillard \& McClintock(2006)]{rm} Remillard, 
R.~A., \& McClintock, J.~E.\ 2006, \araa, 44, 49 

\bibitem[Revnivtsev, Gilfanov \& Churazov (2000)]{rgc00} Revnivtsev, M., Gilfanov, M.,\& Churazov, E. 2000, \aap, 363, 1013

\bibitem[Rhoades \& Ruffini (1974)]{rr74} C. E. Rhoades, C.E. Jr. \&  Ruffini, R., 1974, {\it  Phys. Review Letters}, 32, 324

\bibitem[Rodriguez, Corbel  \& Tomsick (2003)]{rod03} Rodriguez, J., 
Corbel, S., \& Tomsick, J.~A.\ 2003, \apj, 595, 1032 

\bibitem[Rodriguez et al.(2004)]{rod04} Rodriguez, J., 
Corbel, S., Kalemci, E., Tomsick, J.~A., \& Tagger, M.\ 2004, \apj, 612, 
1018 

\bibitem[Rossi et al.(2004)]{rossi04} Rossi, S., Homan, J., 
Miller, J.~M., \& Belloni, T.\ 2004, Nuclear Physics B Proceedings 
Supplements, 132, 416 



\bibitem[Rybicki \& Lightman(1979)]{rl79} Rybicki, G.~B., \& 
Lightman, A.~P.\ 1979, Radiative processes in astrophysics,  New York, Wiley-Interscience   

\bibitem[S{\'a}nchez-Fern{\'a}ndez et 
al.(1999)]{san99} S{\'a}nchez-Fern{\'a}ndez, C., et al.\ 1999, \aap, 348, L9 



\bibitem[Shakura \& Sunyaev  (1973)]{ss73} Shakura, N.I., \& Sunyaev, R.A. 1973, \aap, 24, 337

\bibitem[Shaposhnikov \& Titarchuk (2006)]{ST06} 
Shaposhnikov, N., \& Titarchuk, L. 2006, \apj, 643, 1098 (ST06)

\bibitem[Shaposhnikov \& Titarchuk (2007)]{ST07} 
Shaposhnikov, N., \& Titarchuk, L. 2007, \apj, 663, 445 (ST07)


\bibitem[Shaposhnikov \& Titarchuk(2008)]{st08} Shaposhnikov, N., \& Titarchuk, L.\ 2008, 
AAS/High Energy Astrophysics Division, 10, \#01.08 

\bibitem[Shimura \& Takahara (1995)]{st95}
 Shimura, T., \& Takahara, F. 1995, ApJ, 445, 780 


\bibitem[Shrader \& Titarchuk (2003)]{st03}
 Shrader, C., \& Titarchuk, L.G. 2003,  ApJ, 598, 168 
 
 \bibitem[Shrader \& Titarchuk (1999)]{st99}
 Shrader, C., \& Titarchuk, L.G. 1999,  ApJ, 521, L121 
 
  \bibitem[Sunyaev \& Titarchuk (1980)]{st80}
 Sunyaev, R.A. \& Titarchuk, L.G. 1980,  A\&A, 86, 121 

 
\bibitem[Sobczak et al.(1999)]{sob99} Sobczak, G.~J., 
McClintock, J.~E., Remillard, R.~A., Levine, A.~M., Morgan, E.~H., Bailyn, 
C.~D., \& Orosz, J.~A.\ 1999, \apjl, 517, L121 


\bibitem[Strohmayer et al. (2007)]{str06} Strohmayer, T. E., Mushotzky, R, Winter, L.,  Soria, R.,  Uttley, P.,  Cropper, accepted for publication in ApJ, astro-ph/0701390
 

\bibitem[Titarchuk, \& Fiorito (2004)]{tf04}
Titarchuk, L.G. \& Fiorito, R.  2004,  \apj, 612,  988 (TF04)  

\bibitem[Titarchuk, Lapidus \& Muslimov (1998)]{tlm98}
Titarchuk, L., Lapidus, I.I. \& Muslimov, A. 1998, \apj,  499, 315 (TLM98)


\bibitem[Titarchuk, Mastichiadis \& Kylafis (1997)]{bmc}  Titarchuk, L., Mastichiadis, A., \& Kylafis, N. D., 1997,  \apj, 487, 834


\bibitem[Titarchuk \& Shaposhnikov  (2008)]{ts08}
Titarchuk, L. \&  Shaposhnikov, N.   2008, \apj,  678, 1230




\bibitem[Titarchuk \& Zannias  (1998)]{tz98}
Titarchuk, L. \&  Zannias, T.  1998, \apj,  493, 863

\bibitem[Trudolyubov (2001)]{trud01} Trudolyubov, S.~P.\ 2001, 
\apj, 558, 276 


\bibitem[Trudolyubov,  Borozdin \& Priedhorsky (2001)]{tru01} Trudolyubov, S.~P., 
Borozdin, K.~N., \& Priedhorsky, W.~C.\ 2001, \mnras, 322, 309 


\bibitem[Vignarca et al. (2003)]{vig}
Vignarca, F., Migliari, S., Belloni, T., Psaltis, D., \& van der Klis, M. 2003,  A\&A, 397, 729



\bibitem[Zurita et al.(2002)]{zur02} Zurita, C., et al.\ 
2002, \mnras, 334, 999 

\end{thebibliography}
\end{document}